%% file: main.tex
\newif\ifstandardtemplate\standardtemplatetrue%
\newif\ifelseviertemplate\elseviertemplatefalse%
\newif\ifspringertemplate\springertemplatefalse%
\newif\ifwileytemplate\wileytemplatefalse%
\newcommand{\mytitle}{A variational method for the simulation of hydrogen diffusion in metals}
\newcommand{\myabstract}{We present a new method for the approximate solution of the strongly coupled, nonlinear stress-diffusion problem that appears when modeling hydrogen transport in metals. The most salient feature of the proposed approximation is that it is fully variational, meaning that all the discrete equations are obtained from the optimality conditions of an incremental potential, even for inelastic mechanical behavior. Like other variational methods, the proposed algorithm has remarkable properties, including the symmetry of the tangent operator, making its solution extremely efficient compared to other similar methods available in the literature.}
\newcommand{\myack}{The authors acknowledge the funding received from the Spanish Ministry for Science and Innovation under project EARTH (TED2021-130255B-C32).}

\newcommand{\mypackages}{%
  \usepackage{amssymb}
  \usepackage{amsmath}
  \usepackage{amsthm}
  \usepackage{booktabs}
  \usepackage{caption}
  \usepackage{graphicx}
  \graphicspath{{Figures/}{figures/}}
  \usepackage[dvipsnames]{xcolor}
  \usepackage{todonotes}
  \usepackage{siunitx}
  \usepackage{stackengine}
  \usepackage{subfigure}
  \usepackage{changes}
  \definechangesauthor[name=Ignacio, color=Orange]{IRO}
}

\newcommand{\mymacros}{%
  \newcommand{\concept}[1]{\textbf{\emph{##1}}}
  \newcommand{\defined}{\coloneqq}
  \newcommand{\mbs}[1]{\boldsymbol{##1}}
  \newcommand{\mbsf}[1]{\mbs{\mathsf{##1}}}
  \newcommand{\pairing}[2]{\langle{##1},{##2}\rangle}
  \newcommand{\dd}[2]{\frac{\mathrm{d} ##1}{\mathrm{d} ##2}}
  \newcommand{\pd}[2]{\frac{\partial{##1}}{\partial{##2}}}
  \newcommand{\fd}[2]{\frac{\delta{##1}}{\delta{##2}}}
  \newcommand{\set}[1]{\left\{##1\right\}}
  \newcommand{\trace}{{\mathop{\mathrm{tr}}}}
  \newcommand{\uptohere}{\centerline{\textcolor{blue}{\rule{6cm}{0.2cm}}}}
  \let\oldLambda=\Lambda\renewcommand{\Lambda}{\mathit{\oldLambda}}
  \let\oldGamma=\Gamma\renewcommand{\Gamma}{\mathit{\oldGamma}}
}
\ifstandardtemplate%
\documentclass[10pt,a4paper]{article}
\mypackages%
\mymacros%
\newcommand{\mybibstyle}{unsrt}

\newtheorem{theorem}    {Theorem}[section]

\theoremstyle{definition}

\newtheorem{examplex}[theorem]{$\triangleright\;$Ejemplo}

\theoremstyle{remark}

\title{\mytitle}
\author{E. M. Andr\'es$^{1}$, I. Romero$^{1,2}$}
\date{$^1$Universidad Polit\'ecnica de Madrid,
  Jos\'{e} Guti\'{e}rrez Abascal, 2, 28006 Madrid, Spain\\[2ex]
  $^2$IMDEA Materials Institute,
  Eric Kandel 2, Tecnogetafe\\ 28906 Madrid, Spain
  \\[2ex]\today}

\newenvironment{acknowledgements}{\section*{Acknowledgements}}{}

\begin{document}
\maketitle
\begin{abstract}
  \myabstract%
\end{abstract}
\fi

\ifspringertemplate%
\documentclass[smallcondensed]{svjour3}
\mypackages%
\mymacros%

\title{\mytitle%
\thanks{the thanks}}
\author{\ldots \and Ignacio Romero \and \ldots}
\journalname{The journal name}

\institute{I. Romero \at%
  IMDEA Materials Institute, Eric Kandel 2, Tecnogetafe, Madrid 28906, Spain\\
  Universidad Polit\'ecnica de Madrid, Jos\'e Guti\'errez Abascal, 2, Madrid 29006, Spain\\
  \email{ignacio.romero@imdea.org}
  \and
  XX \at%
  XXX\\
  \email{XXX}}

\titlerunning{\mytitile}
\authorrunning{I. Romero}

\date{Received: date / Accepted: date}

\begin{document}
\maketitle
\begin{abstract}
  \myabstract%
\end{abstract}
\fi

\ifelseviertemplate%
\documentclass[preprint,11pt]{elsarticle}
\mypackages%
\mymacros%
\newcommand{\mybibstyle}{unsrt}

\journal{Computer Methods in Applied Mechanics and Engineering}
\begin{document}
\begin{frontmatter}

\title{\mytitle}

\author[1,2]{Ignacio Romero\corref{cor1}}
\ead{ignacio.romero@upm.es}

\author[3]{xxx}
\ead{xxx}

\address[1]{Universidad Polit\'ecnica de Madrid, Spain}
\address[2]{IMDEA Materials Institute, Spain}
\address[3]{xx}

\cortext[cor1]{Corresponding author. ETS Ingenieros Industriales,
  Jos\'e Guti\'errez Abascal, 2, Madrid 28006, Spain}

\begin{abstract}
\myabstract%
\end{abstract}

\begin{highlights}
\item Research highlight 1
\item Research highlight 2
\end{highlights}

\end{frontmatter}

\fi

\ifwileytemplate%
\documentclass[doublespace,times]{nmeauth}
\mypackages%
\mymacros%
\usepackage{natbib}
\message{Compiling with Wiley template}

\begin{document}
\runningheads{I. Romero}{\dots}

\title{\mytitle}
\author{Ignacio Romero\affil{1}\affil{2}\corrauth}

\address{\affilnum{1} ETSII, Universidad Politécnica de Madrid,
         Jos\'{e} Guti\'{e}rrez Abascal, 2, 28006 Madrid, Spain\break%
         \affilnum{2} IMDEA Materials Institute, Eric Kandel 2, 28096 Getafe, Madrid, Spain}

\corraddr{Dpto.~de Ingenier\'{\i}a Mec\'{a}nica; E.T.S. Ingenieros Industriales;
José Gutiérrez Abascal, 2; 28006 Madrid; Spain. Fax (+34) 91 336 3004}

\begin{abstract}
\end{abstract}

\maketitle
\fi


\input 1intro.tex
\input 2model.tex
\input 3discretization.tex 
\input 4simulations.tex 
\input 5conclusions.tex


\begin{acknowledgements}
  \myack%
\end{acknowledgements}

\bibliographystyle{\mybibstyle}
\bibliography{biblio.bib}

\end{document}

%% file: 1intro.tex
\section{Introduction}
\label{subs-intro}
It has been realized for a long time that the interaction of hydrogen and metals is crucial for determining the embrittlement of the latter and, thus, for the assessment of their strength and reliability \cite{johnson1875br,pfeil1926rd}. These aspects are critical, for instance, in the design of metallic parts of nuclear reactors \cite{takaku1982ad}, pipelines in the oil and gas industry \cite{ohaeri2018pk}, pressure vessels \cite{harries1963xc}, etc. Recently, moreover, the widespread interest in using hydrogen as an energy vector has renewed the efforts in trying to understand this physical phenomenon, its effects on fracture propagation, strength, etc. \cite{hirth1980op,okonkwo2023wp}.

The diffusion of hydrogen through a metal is a complex process that has been thoroughly studied \cite{oriani1994un,aifantis1980be,campbell2009yy}. Hydrogen molecules dissociate at the surface and the hydrogen atoms move through the metallic lattice hopping from one interstitial position of the lattice to another. In the literature, these positions are often referred to as \emph{normal interstitial lattice sites} or \emph{NILS}. A key feature that makes these diffusion problems more challenging is that hydrogen has been shown to accumulate in defects such as vacancies, dislocations, and grain boundaries. Hydrogen at these locations seems to be \emph{trapped} and does not diffuse following the standard kinetics. Moreover, it is now agreed that the trapped hydrogen is the main responsible for the embrittlement of certain metals such as steel. This dual behavior of dissolved hydrogen in metals makes its study challenging and precludes its modeling using simple diffusion theories.

Understanding the materials science aspects of hydrogen diffusion in metals and their embrittlement is just the first step for effectively designing mechanical parts that perform safely. Together with the experimental characterization of transport phenomena, the second necessary step requires incorporating such knowledge into models that can be used to predict part behavior. In most situations, given that existing models are based on complex partial differential equations, this entails the formulation of numerical approximations that are accurate and account for all relevant physical phenomena (see, among many others, \cite{sofronis1989qb,krom1998go,miresmaeili2010lr,dileo2013fu,barrera2016nb,elmukashfi2020vu}). The development of approximations for these \emph{coupled} problems is not exempt from difficulties, especially when incorporating inelastic mechanisms in the solid response and/or temperature. In all cases, moreover, the computational cost entailed in their solution is often quite large.

Motivated by the importance of accurate and efficient numerical methods that model hydrogen diffusion in metals, the current work presents a new discretization method. The method proposed describes the space and time evolution of hydrogen through a metal host, including its trapping and its interaction with the mechanical field. Several chemo-mechanical models for this problem have already been presented in the literature \cite{barrera2016nb,paneda2018yb,elmukashfi2020vu,dinachandra2022cg}. The novelty of the proposed formulation is that we propose a numerical method that is \emph{incrementally variational}, that is, whose governing equations are obtained from the stationarity conditions of a single functional. These equations will encompass the mechanical equilibrium, the diffusion of the free hydrogen as well as the evolution of the trapped hydrogen, all for general inelastic solids.

Incrementally variational methods have been used for more than twenty years now. Initially proposed for quasi-static problems in mechanics \cite{ortiz1999tq}, they have been extended to elastodynamics \cite{kane2000vo}, inelasticity \cite{radovitzky1999kc}, thermo-elasticity \cite{yang2006vg}, thermo-chemo-elasticity \cite{romero2021dd}, topology optimization \cite{bartels2021sz}, phase evolution \cite{carstensen2002uh}, structural models \cite{portillo2020jk}, etc. They have a theoretical appeal because they provide a mathematical framework from where some analysis can be performed. In more practical terms, they furnish a natural setting for adaptivity, one that is far from obvious for transient, inelastic, or coupled problems. Finally, their most pragmatic advantage is that they guarantee the symmetry of the algorithmic tangent, even for coupled inelastic problems. Given that coupled problems invariably entail the solution of large systems of equations, variational updates (asymptotically) halve the computational storage and computational time for direct solvers when compared to standard solvers. These last properties by themselves justify the formulation of variational methods and, in particular, the ones presented herein.

An outline of the remainder of this article follows. In Section~\ref{sec-model}, we review the initial boundary-value problem of stress diffusion in inelastic solids under the assumption of small strain kinematics. There, the balance equations of this problem are recalled, and the thermodynamic foundations are summarized. Section~\ref{sec_variational_formulation} presents the space and time discretization of the problem. Since the spatial discretization is standard, the focus will be on the time discretization. An incremental variational functional will be proposed and it will be shown that its optimality conditions correspond to a time discretization of the coupled problem. Its application to several examples in coupled stress/diffusion of elastic and inelastic solids will be presented in Section \ref{sec_simulations}. Finally, Section~\ref{sec-conclusion} will close the article with a summary of the main findings.


%% file: 2model.tex
\section{A model for stress-diffusion with trapped species}
\label{sec-model}
In this section, we describe a complete model for strongly coupled stress-diffusion in solids with small strain kinematics and general inelastic constitutive response. In addition, the transport equations will track the evolution of \emph{free} as well as \emph{trapped} species. We note that the coupled problem lacks a variational structure; only when we introduce a time discretization we will be able to unveil one, and later exploit it for the formulation of numerical methods. We refer to other references for complete accounts of the theory of diffusion in solids \cite{aifantis1980be} and, more specifically, hydrogen transport in solids \cite{sofronis1989qb,oriani1994un}.

\subsection{Thermodynamics}
\label{subs-thermo}
Before stating the governing equations of the stress-diffusion coupled problem we review the fundamental concepts that stem from the thermodynamics of such systems. The first ingredient for such a description is the identification of the variables that uniquely identify the state of a point in a body that can be subject to inelastic mechanical processes as well as the transport of two species. Borrowing the ideas from equilibrium thermodynamics \cite{callen1985wt}, we assume that the thermodynamic state of such a point is fully described by the array $(\mbs{\varepsilon},\vartheta,\mbs{\xi},\chi_L,\chi_T)$. The first of these variables corresponds to the infinitesimal strain tensor. The second one is the absolute temperature. The third, a set of \emph{internal} variables that are selected to model the inelastic effects of the material and might include, for example, the plastic strain, the damage variable, etc. The symbols $\chi_L$ and $\chi_T$ are employed to denote the concentrations of the two species under consideration. The characters `L' and `T' will later refer to the \emph{lattice} and \emph{trapped} hydrogen; at this point, however, the two species can be taken to be completely independent of each other.

The thermodynamics of the point under consideration is completely described with a potential such as the Helmholtz free energy, a function of the form
\begin{equation}
  \label{eq-helmholtz}
  \psi = \hat{\psi}(\mbs{\varepsilon},\vartheta,\mbs{\xi},\chi_L,\chi_T)
\end{equation}
that encapsulates all the equilibrium features of a representative volume element. In what follows, the temperature will be assumed to be constant and equal to a reference value $\vartheta=\vartheta_0$ and removed from the functional dependency of the free energy and other potentials.

Standard arguments show that the stress tensor $\mbs{\sigma}$, the dissipative forces $\mbs{q}$, and the chemical potentials $\mu_L,\mu_T$ are obtained through the relations
\begin{equation}
  \label{eq-conjugate}
  \mbs{\sigma} = \pd{\psi}{\mbs{\varepsilon}}\ ,\qquad
  \mbs{q} = -\pd{\psi}{\mbs{\xi}}\ ,\qquad
  \mu_L = \pd{\psi}{\chi_L}\ ,\qquad
  \mu_T = \pd{\psi}{\chi_T}\ .
\end{equation}

For future reference, we recall that other thermodynamic potentials can be defined in terms of Legendre transforms of the Helmholtz free energy. For example, the grand potential $\mathcal{G}$ is a function
\begin{equation}
  \label{eq-grand}
  \mathcal{G} = \hat{\mathcal{G}}(\mbs{\varepsilon},\mbs{\xi},\mu_L,\mu_T)
  =
  \sup_{\chi_L,\chi_T}[\hat{\psi}(\mbs{\varepsilon},\mbs{\xi},\chi_L,\chi_T) - \chi_L\,\mu_L - \chi_T\,\mu_T]
\end{equation}
that satisfies
\begin{equation}
  \label{eq-grand2}
  \chi_L = - \pd{\mathcal{G}}{\mu_L}\ ,\qquad
  \chi_T = - \pd{\mathcal{G}}{\mu_T}\ .
\end{equation}

The second derivatives of the thermodynamic potential also have an important physical interpretation. In
particular, the \emph{chemical capacities} are defined as
\begin{equation}
  \label{eq-chem-capacities}
  c_L = \pd{\chi_L}{\mu_L} = - \frac{\partial^2\mathcal{G}}{\partial \mu_L^2}\ ,
  \qquad
  c_T = \pd{\chi_T}{\mu_T} = - \frac{\partial^2\mathcal{G}}{\partial \mu_T^2}\ .
\end{equation}

\subsection{Balance equations}
\label{subs-setting}
We describe first the general setting for stress-diffusion that applies to every possible body with two transported species in chemical equilibrium. The only constraint that will be assumed in this section is that small strain kinematics will suffice to model the mechanical aspects of the solid. Fairly mode complete descriptions can be found in textbooks such as \cite{gurtin2010wu}.

We study a solid that is represented by a domain $\Omega\subset \mathbb{R}^d$, with $d=2$ or $3$. Points in $\Omega$ will be denoted as $\mbs{x}$ and the boundary of this set is $\partial\Omega$. To allow the definition of boundary conditions, we partition $\partial\Omega$ in two different ways
\begin{equation}
  \label{eq-boundary-split}
  \partial\Omega =
  \partial_u\Omega \cup \partial_t\Omega =
  \partial_{\mu}\Omega \cup \partial_j\Omega\ ,
\end{equation}
where
\begin{equation}
  \label{eq-boundary-split2}
  \emptyset =
  \partial_u\Omega \cap \partial_t\Omega =
  \partial_{\mu}\Omega \cap \partial_j\Omega\ .
\end{equation}
Assuming quasistatic deformations, the boundary value problem of mechanics for the displacement field $\mbs{u}$ can be written as
\begin{subequations}
  \label{eq-mechanical}
  \begin{align}
    \nabla\cdot\mbs{\sigma} + \mbs{f} &= 0
    &&\text{in }\Omega\ ,    \label{eq-mech1} \\
    \mbs{\varepsilon} &= \nabla^s\mbs{u}
&&\text{in }\Omega\ , \label{eq-mech11} \\
   \mbs{u} &= \mbs{0} &&\text{on }\partial_u\Omega\ ,\label{eq-mech2} \\
   \mbs{\sigma}\mbs{n} &= \mbs{t} &&\text{on }\partial_t{\Omega}\ ,\label{eq-mech3}
  \end{align}
\end{subequations}
where $\mbs{\sigma}$ is the stress tensor, $\mbs{f}$ is a vector field of volumetric forces,
$\nabla\cdot$ and $\nabla^s$ are, respectively, the divergence and symmetric gradient operators, $\mbs{\varepsilon}$ is the infinitesimal strain tensor, $\mbs{n}$ is the unit vector normal to $\partial_t\Omega$, and $\mbs{t}$ is another vector field of known surface tractions.

Next, we depart from the general transport model discussed heretofore and restrict the balance of mass to a species whose concentration is denoted as $\chi$ and has the peculiarity that it can exist in one of two forms, identified as before with the symbols `$L$' and `$T$'. Hence we write
\begin{equation}
  \label{eq-density-split}
  \chi = \chi_L + \chi_T\ .
\end{equation}
This notation anticipates the study of mass transport in a host where the diffusive species can move through the host lattice, hence the `$L$' symbol, or bind to defects remaining trapped and explaining the symbol `$T$' in the second concentration. These two phases can transform into one another and the key equilibrium condition is that their corresponding chemical potentials must be equal at all points and instants \cite{oriani1994un}, that is,
\begin{equation}
  \label{eq-equilibrium}
  \pd{\psi}{\chi_T} = \pd{\psi}{\chi_L}\ .
\end{equation}
In contrast with the mechanical equilibrium equation which was described by a quasi-static model, we would like to consider, in what follows, the transient behavior of the species' transport. Choosing the mechanical response to be quasi-static makes sense in a stress-diffusion problem because the characteristic times of the mechanical problem are much shorter than the diffusive times. Hence, when compared with the transient diffusive behavior, the mechanical response appears to be \emph{instantaneous}.

To account for mass transport, we define $\mbs{j}_L$ and $\mbs{j}_T$ to be the mass flux vectors of the \emph{lattice} and \emph{trapped} species. Hence, we can define
\begin{equation}
  \label{eq-total-flux}
  \mbs{j} = \mbs{j}_L + \mbs{j}_T
\end{equation}
to be the total mass flux. Using this notation, the mass transport problem may be described by the following initial boundary value problem:
\begin{subequations}
  \label{eq-mt}
  \begin{align}
    \pd{\chi}{t} + \nabla\cdot\mbs{j} &= 0  &&\text{in }\Omega\label{eq-mt-1} \\
    \mu &= \bar{\mu} &&\text{on }\partial_{\mu}\Omega   \label{eq-mt-2}\\
    -\mbs{j}\cdot \mbs{n} &= \bar{\jmath} && \text{on } \partial_j\Omega \label{eq-mt-3}\\
    \chi &= \chi_L + \chi_T \label{eq-mt-4} \\
    \mu_{L} &= \mu_T = \mu \label{eq-mt-5} \\
    \mu(\cdot,0) &= \mu_0\ .
  \end{align}
\end{subequations}

\subsection{Constitutive modeling}
\label{subs-constitutive}
The stress/diffusion problem described by Eqs.~\eqref{eq-mechanical} and~\eqref{eq-mt} is incomplete. To close it, we must complement the thermodynamic potential $\psi$ with kinetic equations for the internal variables and the mass fluxes. These relations cannot be arbitrary since the coupled problem must be such that the total dissipation in an arbitrary process must be non-negative. To enforce this condition, let us first obtain the expression for the dissipation so that it can guide the formulation of correct models.

To start, let us consider the power expended on an arbitrary region $U\subseteq\Omega$. Assuming that there is no volumetric mass supply, this power has the form
\begin{equation}
  \label{eq-power-expended}
  \mathcal{P} =
  \int_U \mbs{f}\cdot \dot{\mbs{u}} \; \mathrm{d} V
  +
  \int_{\partial U} \left((\mbs{\sigma}\mbs{n})\cdot \dot{\mbs{u}} - \mu\,\mbs{j}\cdot \mbs{n}\right) \; \mathrm{d} A\ ,
\end{equation}
where $\mbs{n}$ is the unit normal to the boundary $\partial U$.
The dissipation, the fraction of the expended power that is not transformed into free energy, must be non-negative, i.e.,
\begin{equation}
  \label{eq-dissip}
  \mathcal{D} =
  \mathcal{P} - \dd{}{t} \int_U \hat{\psi}(\mbs{\epsilon}[\mbs{u}],\mbs{\xi},\chi_L,\chi_T) \; \mathrm{d} V\ .
\end{equation}
Employing the divergence theorem on the boundary terms, the balance equations~\eqref{eq-mechanical} and~\eqref{eq-mt}, and the thermodynamic relations~\eqref{eq-conjugate}, we obtain
\begin{equation}
  \label{eq-dissip2}
  \mathcal{D} = \int_U \left( -\mbs{j}\cdot\nabla\mu + \mbs{q}\cdot \dot{\mbs{\xi}} \right) \; \mathrm{d} V\ .
\end{equation}
If we insist that the dissipation must be non-negative in every process and for all subsets $U$, the integrand
of~\eqref{eq-dissip2} must be itself non-negative. To ensure this condition, a standard procedure consists in
formulating a dissipation potential $\Sigma = \hat{\Sigma}(\mbs{q}, \nabla{\mu}_L,\nabla\mu_T; \mbs{\Xi})$, convex in its first argument and concave in the second and third arguments such that $\hat{\Sigma}(\mbs{0},\mbs{0},\mbs{0};\mbs{\Xi})=0$ and
\begin{equation}
  \label{eq-kinetic}
  \dot{\mbs{\xi}} = \pd{\Sigma}{\mbs{q}}\ ,
  \qquad
  \mbs{j}_L = \pd{\Sigma}{\nabla\mu_L}\ ,
  \qquad
  \mbs{j}_T = \pd{\Sigma}{\nabla\mu_T}\ .
\end{equation}
Note that for non-smooth problems such as plasticity, the previous formalism still holds. In this situation, the first kinetic relation must be understood as a relation between the rate of the internal variables and the (convex) subdifferential of~$\Sigma$. We also recall that the \emph{mobility tensors} are defined as
\begin{equation}
  \label{eq-mobility}
  \mbs{m}_{L} = - \pd{\mbs{j}_L}{\nabla\mu_L} =
  -\frac{\partial^2\Sigma}{\partial(\nabla\mu_{L})^2}\ ,
  \qquad
  \mbs{m}_{T} = - \pd{\mbs{j}_T}{\nabla\mu_T} =
  -\frac{\partial^2\Sigma}{\partial(\nabla\mu_{T})^2}\ .
\end{equation}
These definitions allow different mobilities for the species `$L$' and `$T$'. In the next section, for example, we will select a model in which the mobility of the `$T$' species is exactly zero, introducing a simple mechanism to fix the trapped hydrogen.

Another critical feature of this formulation is that the kinetic potential might depend on the state $\mbs{\Xi}$. For instance, later it will be desirable to employ models where the plastic slip rate depends on the concentration of the transported species. The abstract thermodynamic and kinetic models that we have defined heretofore are general enough to accommodate these coupling effects, both in the equilibrium properties as well as in the dissipation processes.

\subsection{A specific model for stress-diffusion of hydrogen in metals}
\label{subs-hydrogen}
In preparation for the numerical examples of Section~\ref{sec_simulations}, and as an application of the ideas of Section~\ref{subs-constitutive}, we present now a complete constitutive model for coupled stress-diffusion of hydrogen in a metal host. The balance equations that describe this problem are precisely the ones introduced in the current section and, as explained before, we close the problem by providing a free energy~$\psi$ and a convex dissipation potential~$\Sigma$. By construction, the resulting model will produce non-negative dissipation in all regions of the solid, and for all processes.

The first ingredient of the model is the Helmholtz free energy which must account for mechanical, chemical, and coupling contributions. Following the work of Di Leo \cite{dileo2013fu}, we assume from the outset the simple split form
\begin{equation}
  \label{eq-helm}
  \psi = \hat{\psi}(\mbs{\varepsilon},\vartheta,\mbs{\xi},\chi_L,\chi_T) =
  \hat{\psi}^{mec}(\mbs{\varepsilon},\mbs{\xi}) + \hat{\psi}^L(\chi_L) + \hat{\psi}^T(\chi_T) + \hat{\psi}^c(\mbs{\varepsilon},\chi_L,\chi_T)\ .
\end{equation}
The first term, $\hat{\psi}^{mec}$ accounts for all the elastic and inelastic mechanical terms; the functions $\hat{\psi}^L,\hat{\psi}^{T}$ incorporate the thermodynamics of the chemical species and are assumed, from the outset, to split additively; finally, the coupling between mechanical and chemical effects is added through the function $\hat{\psi}^c$.

A mechanical free energy will be considered only for $J_2$ elasto-plastic materials. For these, the internal variables will consist of the plastic strain $\mbs{e}^p$, a deviatoric tensor, the strain-like (scalar) isotropic hardening $\alpha$ and the strain-like deviatoric kinematic hardening tensor $\mbs{\beta}$. If we consider, for simplicity, that the isotropic and kinematic hardening are linear, then we would have
\begin{equation}
  \label{eq-psi-mec}
  \hat{\psi}^{mec}(\mbs{\varepsilon},\mbs{\xi})=
  \frac{\kappa}{2} \trace[\mbs{\varepsilon}]^2 
  + 
  G \, \|\mathrm{dev}[\mbs{\varepsilon}]-\mbs{e}^p\|^2
  +
  \frac{H_{iso}}{2} \alpha^2
  +
  \frac{H_{kin}}{3} \|\mbs{\beta}\|^{2}\ ,
\end{equation}
where $\mbs{\xi}$, as indicated before, is the array $(\mbs{e}^p,\alpha,\mbs{\beta})$.
In this energy, $\kappa,G,H_{iso},H_{kin}$ refer to the bulk and shear moduli and the isotropic and kinematic hardening parameters, respectively.

The contribution of the chemical species to the free energy consists of two functions that can be written as
\begin{equation}
  \hat{\psi}^{\alpha} (\chi_{\alpha})  = \mu_{\alpha}^0 \chi_{\alpha} + R \vartheta_0 N_{\alpha} [\theta_{\alpha} \log \theta_{\alpha} + ( 1-\theta_{\alpha}) \log ( 1 - \theta_{\alpha} ) ],
\end{equation}
with $\alpha$ being either $L$ or $T$, representing respectively the ``lattice'' and ``trapped'' contributions,  $\theta_{\alpha} = \chi_{\alpha}/N_{\alpha}$ is the occupancy factor, and $N_{\alpha}$ is the number of atoms of the $\alpha$
species per unit volume of the host. It must be noted that experimental measurements indicate that the trap density is a function of the history of the prior plastic deformation,
\begin{equation}
  N_T = \hat{N}_T (\bar{\mbs{e}}^p).
  \label{nt_dependency_eq}
\end{equation}
In accordance with Sofronis  and McMeeking \cite{sofronis1989qb}, $N_T$ is assumed to increase with the equivalent plastic strain $\bar{\mbs{e}}^p$ according to
\begin{equation}
  N_T = N_{TA} \exp \left[N_{T0} - N_{T1} \exp (N_{T2} \bar{\mbs{e}}^p) \right].
  \label{nt_expression_eq}
\end{equation}
To conclude the definition of all the equilibrium quantities, it remains to define the coupling free energy. For that, we will adopt the usual assumption that the concentrations are coupled only with the \emph{volumetric} part of the strain. For concreteness, we will use
\begin{equation}
  \label{eq-coup-pot}
  \psi^c(\mbs{\varepsilon},\chi_L,\chi_T) = -3\,\kappa\,\gamma\, \trace[\mbs{\varepsilon}](\chi_L+\chi_T-\chi_{L}^0-\chi_{T}^0)\ .
\end{equation}
Here, the constant $\gamma$ denotes the coefficient of chemical expansion and $\chi_{L}^0,\chi_{T}^0$, two reference concentration values of the two species. We note that, in this simple model, the inclusion of either of the two species can only induce isotropic, volumetric expansion and this expansion does not differentiate between the two types of inclusions.

From the thermodynamic definition~\eqref{eq-conjugate}, the chemical potentials of the two species can be found
\begin{equation}
  \mu_{\alpha} = \frac{\partial \psi_{\alpha}}{\partial \chi_{\alpha}}
  =
  \mu_{\alpha}^0 + R \vartheta_0 \log \left( \frac{\theta_{\alpha}}{ 1-\theta_{\alpha}} \right)
- 3\,\kappa\,\gamma\, \trace[\mbs{\varepsilon}]\ ,
  \label{mu_alpha_def_eq}
\end{equation}
and also inverted,
\begin{equation}
  \chi_{\alpha} = \frac{N_{\alpha}E_{\alpha}}{1+E_{\alpha}}
  \ ,
  \qquad
  \mathrm{with}
  \quad
  E_{\alpha} = \exp{\big(\frac{\mu_{\alpha}-\mu_{\alpha}^0+3\,\kappa\,\trace[\mbs{\varepsilon}]}{R\vartheta_0}\big)}.
\end{equation}
The fundamental condition for chemical equilibrium is that, at any point and instant, the
two forms of hydrogen must have identical chemical potentials, i.e., $\mu_L=\mu_T$. This requirement
is often credited to Oriani \cite{oriani1994un} and, using Eq.~\eqref{mu_alpha_def_eq}, it gives
\begin{equation}
  \frac{\theta_{T}}{1 - \theta_T} = \frac{\theta_{L}}{1 - \theta_L} K_T,
  \label{eq-thetas}
\end{equation}
or, equivalently,
\begin{equation}
  \chi_T = \frac{K_T N_T \chi_L}{N_L - \chi_L + K_T \chi_L},
  \label{relation_theta_T_theta_L}
\end{equation}
with
\begin{equation}
  K_T = \exp{\left( \frac{W_B}{R \theta_0} \right)}\ ,
  \qquad
  \mathrm{and}
  \qquad
  W_B = \mu_{L 0} - \mu_{T 0}\ .
\end{equation}

The stress-diffusion model is closed with the definition of the kinetic potential. For simplicity, we will assume that this function splits into a contribution that provides the rates of the mechanical fields and another one that gives the mass flux, that is
\begin{equation}
  \label{eq-h-kinetic}
  \hat{\Sigma}(\mbs{q},\nabla\mu_L,\nabla\mu_T;\mbs{\Xi}) =
  \hat{\Sigma}^{mec}(\mbs{q};\mbs{\Xi}) + \hat{\Sigma}^{chem}(\nabla\mu_L,\nabla\mu_T; \mbs{\Xi})\ .
\end{equation}
The mechanical contribution depends, naturally, on the process that would like to be accounted for. To be specific, in the examples of Section~\ref{sec_simulations} we will consider elastic and elasto-plastic mechanical models. In the first case, there are no internal variables and the mechanical part of the kinetic potential need not be defined. To define the kinetic potential of an elastoplastic material, let us recall that the \emph{elastic domain} of one of these materials is the set of points $\mbs{q}$ such that
\begin{equation}
  \label{eq-yield}
  f(\mbs{q}; \mbs{\Xi}) \le 0\ ,
\end{equation}
where $f$ is the \emph{yield function}. Then, the kinetic potential is the indicator function of the elastic domain, namely,
\begin{equation}
  \label{eq-elastoplastic-k}
  \hat{\Sigma}^{mec}(\mbs{q}; \mbs{\Xi}) =
  \begin{cases}
    0 & \mathrm{if}\ f(\mbs{q}; \mbs{\Xi}) \le 0\ ,\\
    +\infty & \mathrm{otherwise} .
  \end{cases}
\end{equation}
If the function $f$ is convex on its first argument, the elastic domain will be a convex set and the kinetic potential will be convex, albeit non-smooth. Notice that the concentration enters into the definition of the yield function and can modify, for example, the yield stress, coupling the chemical and mechanical fields at the kinetic level.

The chemical contribution to the dissipative potential is taken to be
\begin{equation}
  \label{eq-chemical-k}
  \hat{\Sigma}^{chem}(\nabla\mu_L,\nabla\mu_T; \mbs{\Xi}) = - \frac{m_L(\mbs{\Xi})}{2} |\nabla\mu_L|^2\ .
\end{equation}
Note that, for this choice, the mobility of the \emph{lattice} hydrogen is isotropic and the \emph{trapped} hydrogen can not diffuse. The assumption that the mass flux is proportional to the gradient of the chemical potential is usually accepted when the concentration is low and there are no interactions among the species under consideration. A frequently employed relation selects the mobility to be a function of the concentration.

The model presented in this section describes equilibrium states and kinetics of metal/hydrogen systems where the coupling between the chemical and mechanical fields appears in the coupling contribution of the free energy and the mechanical part of the kinetic potential. More accurate models, for example, may not allow simple splits of the form~\eqref{eq-helm} but demand, rather, tighter couplings. In particular, the model outlined above assumes that the elastic constants and hardening parameters have values that are independent of the hydrogen concentration, which might be reasonable for dilute configurations. The effects of hydrogen on the yield stress of the metal are usually more relevant and are explicitly accounted for in our model through the dependency of $\Sigma^{mec}$ on $\mbs{\Xi}$. Following the ideas presented before, more general coupled models can be easily formulated.

The focus of this article is on the formulation of numerical methods which, to a large extent, is independent of the stress-diffusion constitutive and kinetic model selected. The one described in this section is general and complex enough to illustrate the key features of the new algorithms and thus we refrain from using more elaborate models.


%% file: 3discretization.tex
\section{Variational formulation of the rate problem}
\label{sec_variational_formulation}
The initial boundary value problem of stress-diffusion described in Section~\ref{sec-model} is given by the mechanical and chemical balance laws, the corresponding initial and boundary conditions, together with the constitutive relations and the kinetic relations. This strongly coupled problem lacks a variational structure for the field unknowns. In this section, we will show that when we introduce a \emph{time} discretization, the whole problem can be formulated in such a framework and all the equations of the problem can be recovered as the stationarity conditions of a single incremental functional.

\subsection{Motivation}
The advantages of variational formulations are manifold and have been outlined in Section~\ref{subs-intro}. Expanding on it, we claim that there are three types of advantages in such a formulation. First, from a fundamental point of view, these approximations endow the problem with the structure of a minimization or a saddle point optimization program. This framework opens the door to new error estimators \cite{radovitzky1999kc,portillo2020jk}, special iterative solvers such as the nonlinear conjugate gradient \cite{dai2010yx} and Uzawa's method \cite{hu2006xg}. Second, from a purely computational standpoint, variational methods guarantee the symmetry of the tangent matrix required for nonlinear, Newton-type solutions. This, by itself, halves the storage requirement of such matrices and, asymptotically, the solution time of direct solvers as well. Some of these results will be showcased in Section~\ref{sec_simulations}. Last, the existence of an incremental variational principle can sometimes be exploited to study mathematical aspects of the (incremental) problem such as its existence or the convergence of approximations to the exact solution. We refer to, e.g., \cite{romero2021dd} for a more extensive discussion of these and related aspects.

\subsection{Time discretization}
No variational principle has been found behind the stress-diffusion problem, as described in Section~\ref{sec-model}. We will, however, exploit the ideas of \emph{variational updates} to reveal a variational principle behind the time-discrete version of such a problem.

Before we present this principle, let us introduce $K+1$ instants $0 \equiv t_0 < t_1 < ... < t_k < t_{k+1} < ... < t_K\equiv T$ in the interval $[0, T]$ of analysis. Let $x_k$ denote the approximated value of any variable $x$ at time $t_k$ and $\Delta t_k=t_{k+1}-t_k$. A time marching method, or update, for the problem under consideration is a rule for obtaining the state in the body at time $t_{k+1}$ from data at the previous time step and information of the external stimuli in the time interval $[t_k, t_{k+1}]$.

For the stress-diffusion problem, let us write the complete thermo-chemo-mechanical state of the body at time $t$ as
\begin{equation}
  \label{eq-state}
   \Xi(\mbs{x},t) = (\mbs{u}(\mbs{x},t), \mbs{\xi}(\mbs{x},t), \mu^L(\mbs{x},t), \mu^T(\mbs{x},t))\ ,
\end{equation}
an array that collects the displacement, internal variables, and chemical potentials at every point of the body. Once these fields are known, all the remaining quantities of interest (stresses, concentrations, max fluxes, etc.) can be obtained employing the constitutive models of Section~\ref{subs-constitutive}.

A variational update is a numerical method that,
given $\Xi_k\approx \Xi(\cdot,t_k)$, provides a map $\Xi_k\mapsto\Xi_{k+1}$ obtained from the stationarity condition of a single functional. This map should be consistent with the true evolution problem in the sense that the exact equilibrium and evolution equations of the problem should be recovered from the stationarity condition when $\Delta t_k\to0$. If this is verified, the variational update will yield a sequence of states $(\Xi_{0},\Xi_1,\ldots,\Xi_K)$ that approximates the solution of the stress-diffusion problem.

\subsection{An incremental functional}
\label{subs-incremental}
To formulate the desired incremental functional, we extend our previous work \cite{romero2021dd} to incorporate two transported species under the constraint that their chemical potentials should be equal. Before that, and abusing slightly the notation, let us extend the notion of chemo-mechanical state and redefine
\begin{equation}
  \label{eq-state2}
  \Xi_k =
  (\mbs{u}_k, \mbs{\xi}_k, \mu^L_k, \mu^T_k, \chi^L_k, \chi^T_k,\lambda_{k}
  )\ ,
\end{equation}
where $\lambda_k$ is a new scalar field and ignoring, for the moment, that the concentrations can be obtained from the chemical potentials, the strains, and the internal variables as indicated in Eq.~\eqref{eq-grand2}. Assuming now that the state $\Xi_k$ is given, we introduce the functional
\begin{equation}
  \begin{split}
    \Phi_k[\Xi_{k+1}] =
    & \int_\Omega
      \left[ \hat{\psi}(\mbs{\varepsilon}[\mbs{u}_{k+1}],\mbs{\xi}_{k+1},\chi_{k+1}^L,\chi_{k+1}^T) -
      \psi_k
      \right]
      \; \mathrm{d} V
    \\
    & - \int_{\Omega}
      \left[
      \mu^L_{k+1} ( \chi^L_{k+1} - \chi^L_{n}) +
      \mu^T_{k+1} ( \chi^T_{k+1} - \chi^T_{n})
      \right]
      \; \mathrm{d} V
      \\
    & +
      \int_\Omega
      \Delta t_k \,
      \Sigma
      \left(
      \frac{\mbs{\xi}_{k+1}-\mbs{\xi}_k}{\Delta t_k}, \nabla{\mu}^L_{k+1}; \Xi_k
      \right)
      \; \mathrm{d} V
      \\
    & - \int_\Omega
      \mbs{f} \cdot (\mbs{u}_{k+1} - \mbs{u}_k)
      \; \mathrm{d} V \\
    & - \int_\Omega
      \lambda_{k+1} (\mu^L_{k+1} - \mu^T_{k+1})
      \; \mathrm{d} V
      \\
    & + \int_{\partial_j \Omega}
      \Delta t_k\, \bar{\jmath}\, \mu^L_{k+1}
      \; \mathrm{d} A
      \\
    & - \int_{\partial_t \Omega}
      \mbs{t} \cdot (\mbs{u}_{k+1} - \mbs{u}_k)
      \; \mathrm{d} A .
 \end{split}
 \label{eq-inc}
\end{equation}
To alleviate the notation, we have replaced $\hat{\psi}(\mbs{\varepsilon}[\mbs{u}_{k}],\mbs{\xi}_{k},\chi_{k}^L,\chi_{k}^T)$ in this definition with $\psi_k$, which plays the role of a constant for the potential~$\Phi_k$.

Let us make a few remarks regarding the expression of $\Phi_k$. First, we note that the definition of this potential must be updated in every time step since it depends parametrically on $\Xi_k$. Also, let us note that it incorporates terms that point towards a variational statement of the mechanical as well as the transport problem. In addition to a term that aims to account for the dissipative effects, Oriani's condition is enforced weakly in the fifth integral of Eq.~\eqref{eq-inc}.

We show next that the stationarity conditions of the potential~\eqref{eq-inc} correspond to the \emph{weak} formulation of the balance equations of mass and momentum, the constitutive equations, and the kinetic equations of the coupled problem. For that, it suffices to calculate the variations of $\Phi_k$ with respect to each of the fields in the state $\Xi_{k+1}$ and set them equal to zero. Thus, we define the variational update to be the implicit map that provides $\Xi_{k+1}$ from $\Xi_k$ by solving
\begin{equation}
  \delta\Phi_k(\Xi_{k+1}) =
  D_{\Xi}\Phi_k(\Xi_{k+1}) \cdot \delta\Xi = 0,
  \label{stationary_condition}
\end{equation}
for all admissible variations of the state $\delta \Xi_{k+1}$. Next, we expand the last equation by taking the (partial) variations of the incremental potential with respect to each of the fields in the state, to obtain,
\begin{subequations}
  \label{eq-opti}
  \begin{align}
      D_{\mbs{u}} \Phi_k \cdot \delta \mbs{u}
      = & \int_\Omega
          \left[
          \pd{\psi_{k+1}}{\mbs{\varepsilon}} \cdot \mbs{\varepsilon}[\delta \mbs{u}]
          - \mbs{f} \cdot \delta \mbs{u}
          \right]
          \; \mathrm{d} V
        -
          \int_{\partial_t \Omega} \mbs{t} \cdot \delta \mbs{u}
          \; \mathrm{d} A\ ,
          \label{eq-opti1}
    \\
      D_{\mbs{\xi}} \Phi_k \cdot \delta \mbs{\xi}
      = & \int_\Omega
          \left[
          \pd{\psi_{k+1}}{\mbs{\xi}} \cdot \delta \mbs{\xi}
          +
          \pd{\Sigma}{\mbs{\xi}}\cdot \partial\mbs{\xi}
          \right]
          \; \mathrm{d} V
          \label{eq-opti2}
    \\
	D_{\mu^L} \Phi_k \cdot \delta \mu^L
    = & \int_\Omega
        \left[
        - (\chi^L_{k+1} - \chi^L_k)\; \delta\mu^L
        - \lambda_{k+1} \, \delta \mu^L
        \right]
        \; \mathrm{d} V\ ,
        \nonumber
        \\
        & + \int_\Omega
          \Delta t_k \pd{\Sigma_{k+1}}{\nabla\mu^L} \cdot \nabla (\delta \mu^L)
          \; \mathrm{d} V
         + \int_{\partial_j \Omega} \Delta t_k \, \bar{j}\; \delta \mu^L
          \; \mathrm{d} A\ ,
          \label{eq-opti3}
    \\
    D \Phi_k \cdot \delta \mu^T
    =& \int_\Omega
       \left[
       -(\chi^T_{k+1} - \chi^T_k) \delta \mu^T +
       \lambda_{k+1} \delta \mu^T
       \right]
       \; \mathrm{d} V\ ,
       \label{eq-opti4}
    \\
	D \Phi_k \cdot \delta \chi^L
    = & \int_\Omega
        \left( \pd{\psi_{k+1}}{\chi^L} - \mu^L_{k+1} \right) \delta \chi^L
        \; \mathrm{d} V\ ,
        \label{eq-opti5}
    \\
	D \Phi_k \cdot \delta \chi^T
    = & \int_\Omega
        \left( \pd{\psi_{k+1}}{\chi^T} - \mu^T_{k+1}
        \right) \delta \chi^T
        \; \mathrm{d} V\ ,
        \label{eq-opti6}
    \\
	D \Phi_k \cdot \delta \lambda
      = & \int_\Omega
          -\left( \mu^L_{k+1} - \mu^T_{k+1} \right) \delta \lambda
          \; \mathrm{d} V \ .
          \label{eq-opti7}
  \end{align}
\end{subequations}
For the sake of notational simplicity, the functional dependence of the potentials has been omitted in this set of equations. We remark that Eqs.~\eqref{eq-opti1} and~\eqref{eq-opti3} are the weak statements, respectively, of the balance of momentum and mass. In turn, Eqs. \eqref{eq-opti2}, \eqref{eq-opti5} and \eqref{eq-opti6} correspond, respectively, to the definitions of the thermodynamic fluxes that are conjugate to the internal variables and the two types of concentrations. Finally, Eq. \eqref{eq-opti7} is Oriani's chemical equilibrium condition.

Relations~\eqref{stationary_condition} and~\eqref{eq-opti} indicate that the sought update map $\Xi_k\mapsto\Xi_{k+1}$ is the solution of seven variational equations. However, at least formally, many of the unknowns can be obtained in closed form and eliminated from the solution. First, from Eq.~\eqref{eq-opti7} we recover Oriani's condition $\mu_{k+1}^L=\mu_{k+1}^{T}=\mu_{k+1}$. Then, inverting Eqs.~\eqref{eq-opti5} and~\eqref{eq-opti6} we can obtain the concentrations $\chi_{k+1}^L,\chi_{k+1}^T$ in terms of the reduced set of state variables $(\mbs{\varepsilon}_{k+1},\mbs{\xi}_{k+1}, \mu_{k+1})$. Next, the Lagrange multiplier $\lambda_{k+1}$ is found to be equal to $\chi_{k+1}^T-\chi_k^{T}$ from Eq.~\eqref{eq-opti4}. Finally, Eq.~\eqref{eq-opti2} gives the standard relation for the internal variables as a function of the strain $\varepsilon_{k+1}$ and $\mu_{k+1}$. As a result of all these simplifications, there exists a functional that depends only on \emph{two fields} defined as
\begin{equation}
  \label{eq-two-field}
  \begin{split}
  \Psi_k &=
  \tilde{\Psi}_k[\mbs{u}_{k+1},\mu_{k+1}] \\
	& =
 \min_{\mbs{\xi}_{k+1}, \chi^L_{k+1}, \chi^T_{k+1}}
 \max_{\lambda_{k+1}}
 \Phi (\mbs{\varepsilon_{k+1}}, \mbs{\xi_{k+1}}, \chi^L_{k+1}, \chi^T_{k+1}, \mu_{k+1}, \mu_{k+1}, \lambda_{k+1}) ,
  \end{split}
\end{equation}
whose stationarity provides the update equations. For all practical purposes, this amounts to proving that the variational formulation can be efficiently implemented as a two-field problem, just like a standard (non-variational) formulation of the nonlinear stress/diffusion problem.

\subsection{Spatial discretization}
The ideas described in Section~\ref{subs-incremental} furnish a time discretization of the initial boundary-value problem of coupled stress-diffusion. To complete the discretization required to implement a numerical solution, we describe next the spatial approximation of the equations in the context of the finite element method. Let us stress that the variational update proposed is compatible with any Galerkin-type method, and the one described next is the simplest. It could be extended to other, more sophisticated ones as long as they do not spoil the variational character of the formulation. In particular, even though we restrict our presentation to a ``displacement finite element method'', other variational methods could be introduced without any complications to address, for example, locking problems in the quasi-incompressible case. In particular, any inf-sup stable pressure-displacement formulation could be adopted, as well as mixed methods based on an average treatment of the volumetric strain such as the B-bar, enhanced, or incompatible modes methods \cite{simo1990uj,brezzi1991tn,bischoff2004,romero-bubbles-2007}.

As advanced, to close the discretization of the stress-diffusion problem, consider a partition of the domain $\Omega$ into a collection of compatible and disjoint elements $\Omega_e$ with $e=1,\ldots,n_E$ connected at $n_N$ nodes with coordinates $\mbs{x}_a,\; a=1,\ldots,n_N$. Associated with this finite element mesh, nodal piecewise polynomial shape functions $N^a:\Omega\to\mathbb{R}$ are defined satisfying $N^a(\mbs{x}_b) = \delta^a_b$, where $\delta^a_b$ is the Kronecker delta. The usual Galerkin \emph{ansatz} starts by defining the unknown fields at time $t_k$ as linear combinations of the shape functions, i.e.,
\begin{equation}
  \label{eq-sp-discrete}
  \begin{split}
    \mbs{u}^h_k(\mbs{x}) &= \sum_{a=1}^{n_N} N^a(\mbs{x})\; \mbs{u}_k^a\ ,\\
    \mu^h_k(\mbs{x}) &= \sum_{a=1}^{n_N} N^a(\mbs{x})\; {\mu}_k^a\ ,
  \end{split}
\end{equation}
with the nodal values of the displacement and chemical potential on node $a$ at time $t_k$ denoted respectively as $\mbs{u}_k^a$ and $\mu^a_k$.
The nodal values of the displacement and the chemical potential must be fixed at the corresponding Dirichlet boundary, that is, $\partial_u\Omega$ and $\partial_{\mu}\Omega$. With this notation, we close the space and time discretization summarizing that, given some approximate solution $\mbs{u}^h_k,\mu^h_k$ at time $t_k$, the solution at time $t_{k+1}$ would be obtained by optimizing
\begin{equation}
  \label{eq-final}
  \begin{split}
    \inf_{\mbs{u}^h_{k+1},\mu^h_{k+1}}& \tilde{\Psi}_k (\mbs{u}^h_{k+1},\mu^h_{k+1}) \\
    &=
  \inf_{\mbs{u}^h_{k+1},\mu^h_{k+1},\mbs{\xi}_{k+1},\chi^L_{k+1},\chi_{k+1}^T} \sup_{\lambda_{k+1}}
    \Psi_k (\mbs{u}^h_{k+1},\mu^h_{k+1}, \mbs{\xi}_{k+1}, \chi^L_{k+1},\chi_{k+1}^T,\lambda_{k+1})\ .
    \end{split}
  \end{equation}

%

%% file: 4simulations.tex
\section{Numerical simulations}
\label{sec_simulations}
In this section, we show three simulations that illustrate the solution of stress-diffusion coupled problems of the type described in Section~\ref{sec-model} using the methods introduced in Section~\ref{sec_variational_formulation}. They have been chosen with goals in mind. On the one hand, we would like to show the generality of the variational method proposed. On the other hand, we aim to emphasize certain features of variational updates, such as their relation with an incremental minimization functional and the benefits of such a choice. We note that we do not intend to present a thorough analysis of the specific mechanical problems selected.

All simulations have been run in IRIS, our in-house finite element code. For some of the results, we have used as a reference the standard finite element formulation of the coupled problem where each of the two balance equations is discretized independently, without any link to a variational principle. For this implementation, the displacement vector field and the equilibrium chemical potential are selected as independent fields so that the number of unknowns is the same as in the variational method.

For the standard formulation of the problem, we have used shared-memory versions of the Pardiso direct linear solver \cite{petra2014op}. For the variational method, we have employed a solver that exploits the full symmetry of the tangent.

\subsection{Elasticity coupled with hydrogen diffusion}
\label{example-elastic}
In this first example, we show the numerical solution of a small strain elastic solid strongly coupled with hydrogen diffusion. For that, we consider a cylinder of radius 0.5~mm and height 1~mm of a linear elastic material with the parameters given in Table~\ref{table_mat_parameters_elastic}.

\begin{table}[h]
	\caption{Material parameters for example~\ref{example-elastic}.}
	\centering
    \renewcommand{\arraystretch}{1.2}
	\begin{tabular}{l c c c c}
	\toprule
	Parameter & Symbol & Value & Units\\
	\hline
    Young's modulus & $E$ & $2.07\cdot 10^5$ & MPa\\
	Poisson's ratio & $\nu$ & $0.3$ & - \\
	Bulk modulus & $\kappa$ & $1.73\cdot10^5$ & MPa\\
	Shear modulus & $G$ & $7.96\cdot10^4$  & MPa\\
	Chemical expansion coefficient & $\gamma$ & $6.67 \cdot10^{4}$ & mm$^3$/mol\\
	Hydrogen diffusivity & $D$ & $1.27\cdot 10^{-2}$  & mm$^2$/s \\
	Baseline chemical potential & $\mu_{0}$ & $-19.58 \cdot 10^7$  & mJ/mol\\
        Ref. chemical potential $L$ & $\mu_{L}^0$ & $2.86 \cdot 10^7$  & mJ/mol\\
	Ref. chemical potential $T$ & $\mu_{T}^0$ & $-3.14 \cdot 10^7$  & mJ/mol\\
	Density of $L$ atoms & $N_L$ & $8.47 \cdot 10^{-4}$  & mol/mm$^3$\\
	First parameter for density of $T$ atoms & $N_{T0}$ & $2.33 \cdot 10^1$ & -\\
 Second parameter for density of $T$ atoms & $N_{T1}$ & $2.33$ & -\\
 Third parameter for density of $T$ atoms & $N_{T2}$ & $-5.50$ & -\\
 Fourth parameter for density of $T$ atoms & $N_{TA}$ & $1.66 \cdot 10^{-33}$ & mol/mm$^3$\\
	Universal gas constant & $R$ & $8.31 \cdot 10^{3}$  & mJ/K\,mol\\
	Ref. temperature & $\theta_0$ & 300  & K\\ 
	\bottomrule
	\end{tabular}
	\label{table_mat_parameters_elastic}
\end{table}
The cylindrical solid is subject to a uniform pressure which is ramped up to a maximum value of $100$ MPa at a constant rate during 1~s on the central circular surface of radius 0.15~mm on the upper flat face, while precluding any displacement and any change in the chemical potential on the bottom flat surface.

For the numerical model, only one fourth of the billet is modeled. On the symmetry surface of the $xz$ and $yz$ planes, the displacements in the y-direction and x-direction, respectively, are set to zero. For this part, we use 63037 tetrahedral elements with small strain kinematics. A standard formulation is employed with full integration. We will compare the solutions obtained by a standard discretization of the problem using the backward Euler integration scheme in time, vs. the variational solution. In both cases, the simulation is performed with a constant time step size equal to 0.01~s.

The mechanical material model is based on a standard linear elastic model. The free energy of the material and the rest of the model are detailed in Section~\ref{subs-hydrogen}.

\begin{figure}
	\centering
	\includegraphics[width=0.45\textwidth]{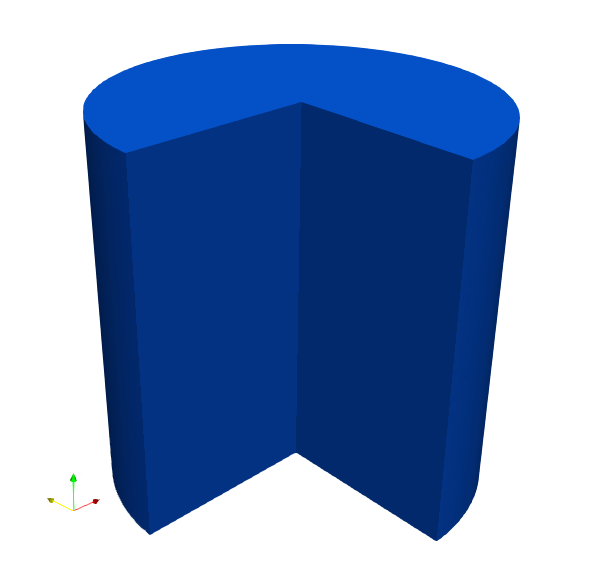}
	\includegraphics[width=0.45\textwidth]{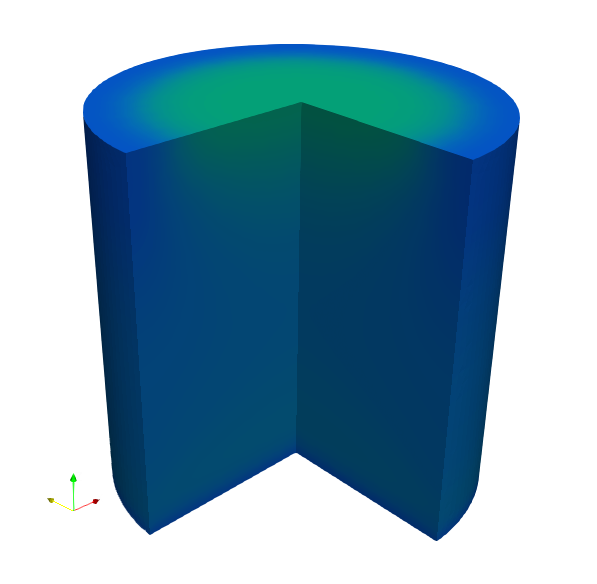}\\
	(a) $t = 0.0$ s $\hspace{3.75 cm}$ (b) $t = 0.2$ s\\
	\includegraphics[width=0.45\textwidth]{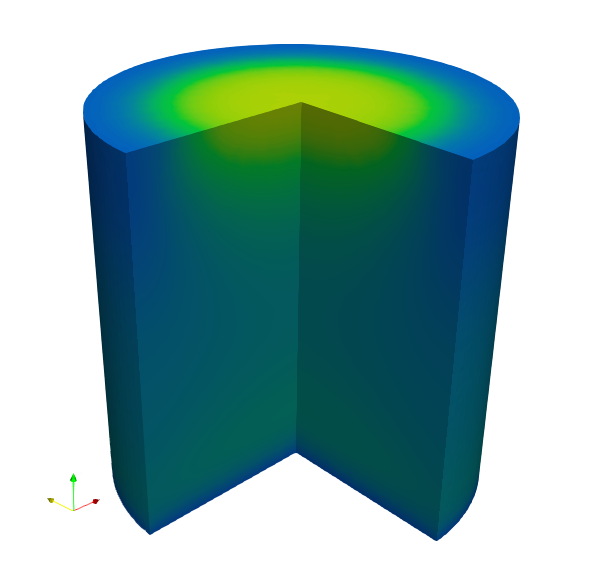}
	\includegraphics[width=0.45\textwidth]{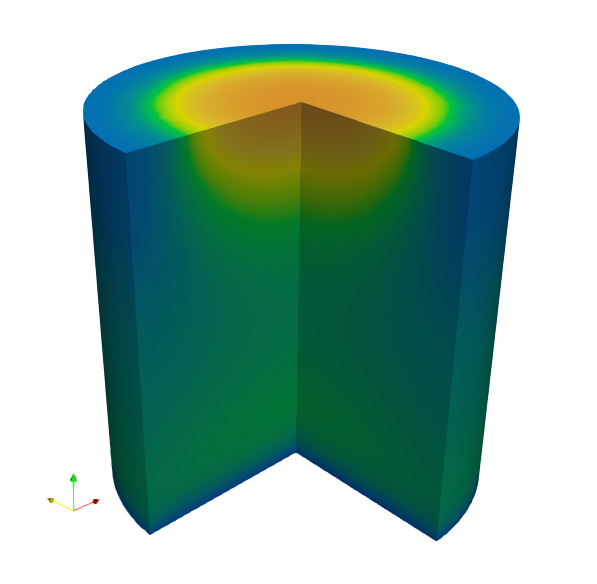}\\
	(c) $t = 0.4$ s $\hspace{3.75 cm}$ (d) $t = 0.6$ s\\
	\includegraphics[width=0.45\textwidth]{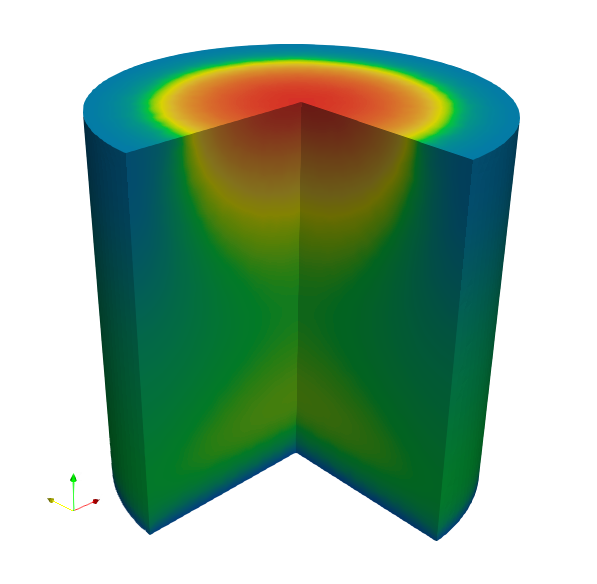}
	\includegraphics[width=0.45\textwidth]{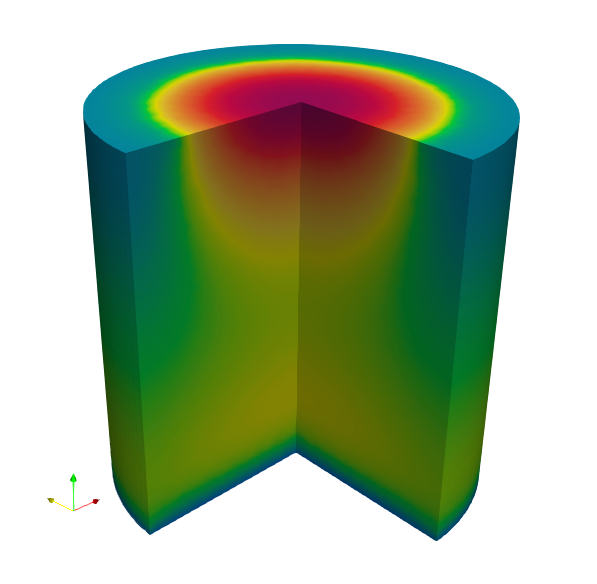} \\
	(e) $t = 0.8$ s $\hspace{3.75 cm}$ (f) $t = 1.0$ s\\
	\includegraphics[width=0.8\textwidth]{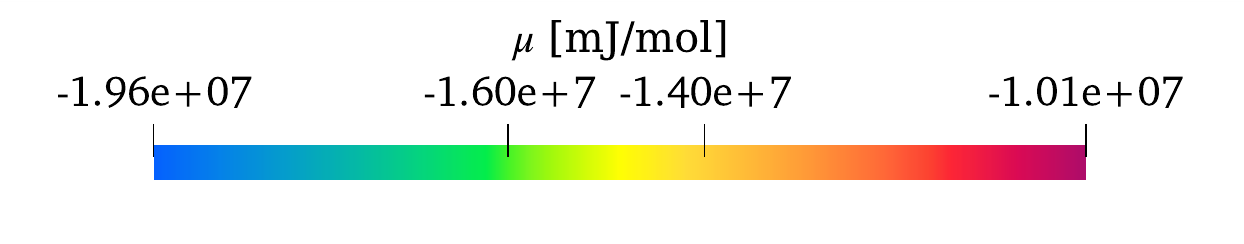}\\
	\caption{Example~\ref{example-elastic}: Chemical potential at different time instants.}
	\label{fig_mu_elastic}
\end{figure}

\begin{figure}
	\centering
	\includegraphics[width=0.45\textwidth]{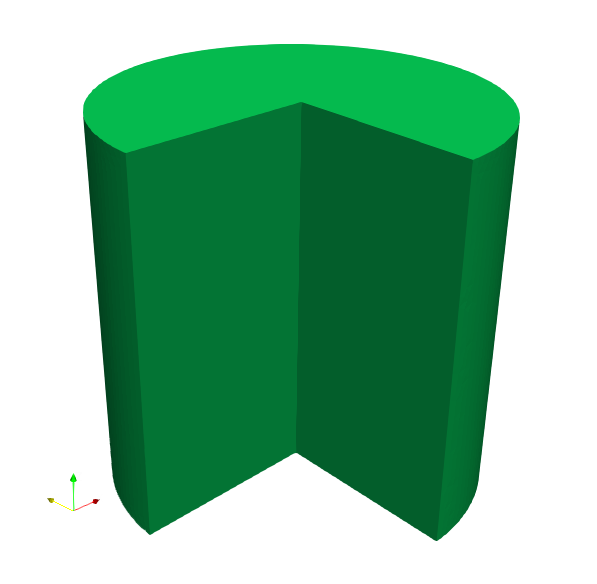}
	\includegraphics[width=0.45\textwidth]{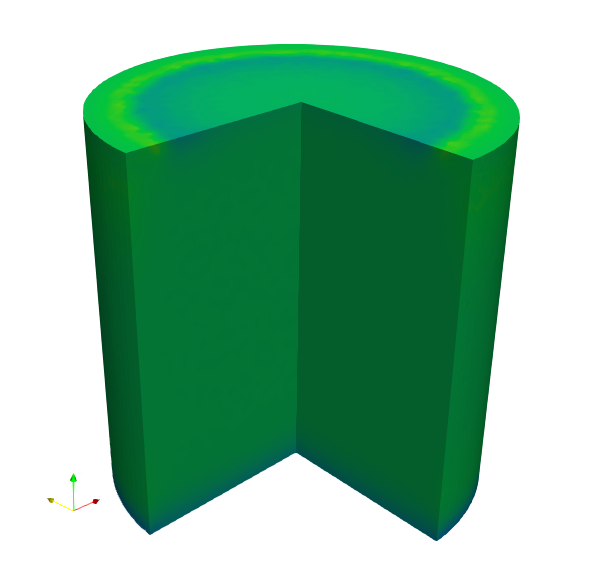}\\
	(a) $t = 0.0$ s $\hspace{3.75 cm}$ (b) $t = 0.2$ s\\
	\includegraphics[width=0.45\textwidth]{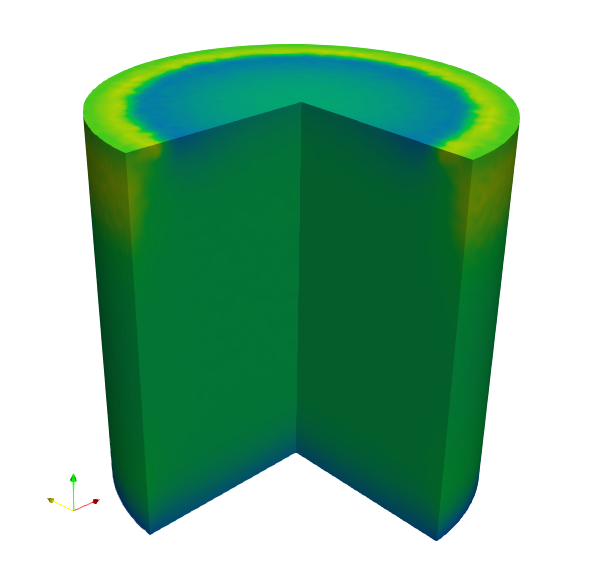}
	\includegraphics[width=0.45\textwidth]{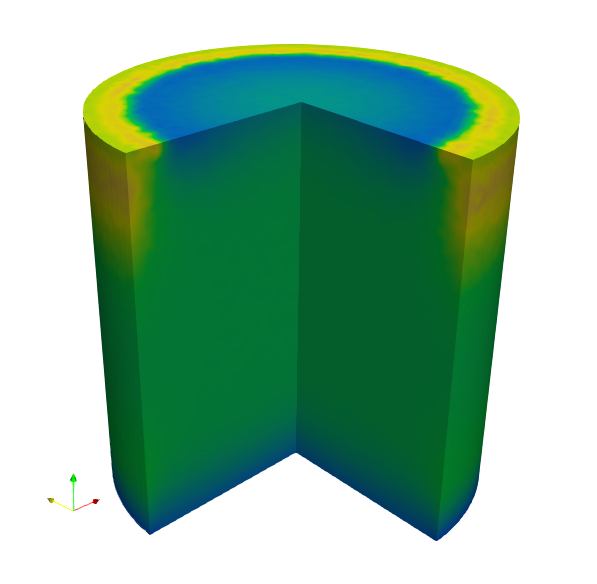}\\
	(c) $t = 0.4$ s $\hspace{3.75 cm}$ (d) $t = 0.6$ s\\
	\includegraphics[width=0.45\textwidth]{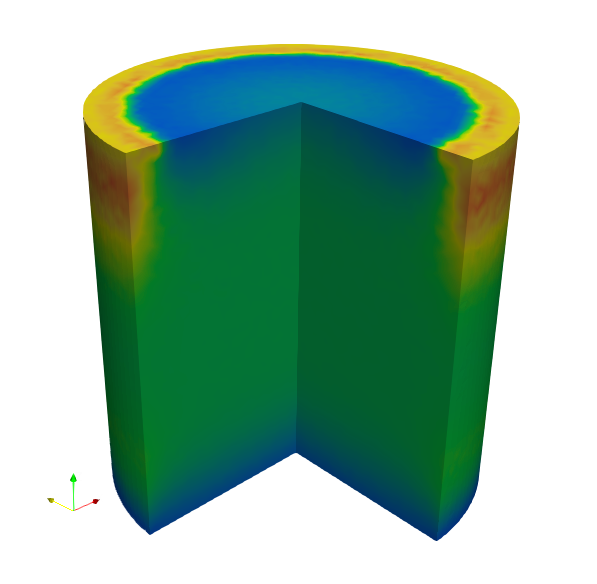}
	\includegraphics[width=0.45\textwidth]{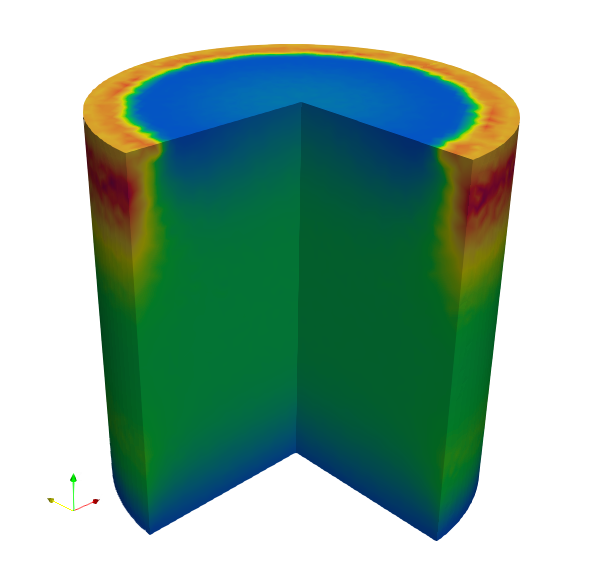}\\
	(e) $t = 0.8$ s $\hspace{3.75 cm}$ (f) $t = 1.0$ s\\
	\includegraphics[width=0.8\textwidth]{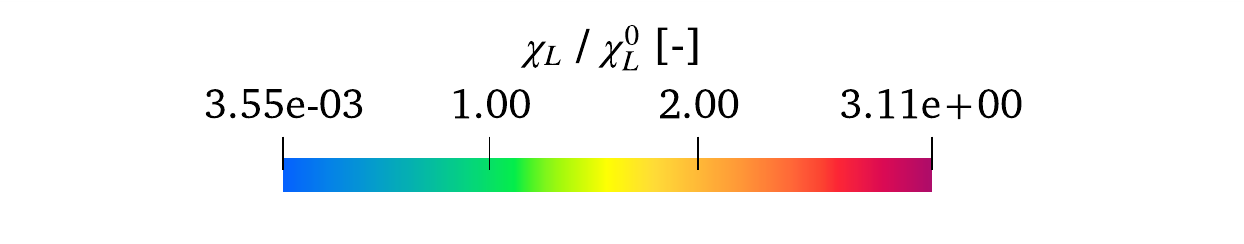}||
	\caption{Example~\ref{example-elastic}: Normalized lattice hydrogen concentration at different time instants.}
	\label{fig_chiL_elastic}
\end{figure}

\begin{figure}
	\centering
	\includegraphics[width=0.45\textwidth]{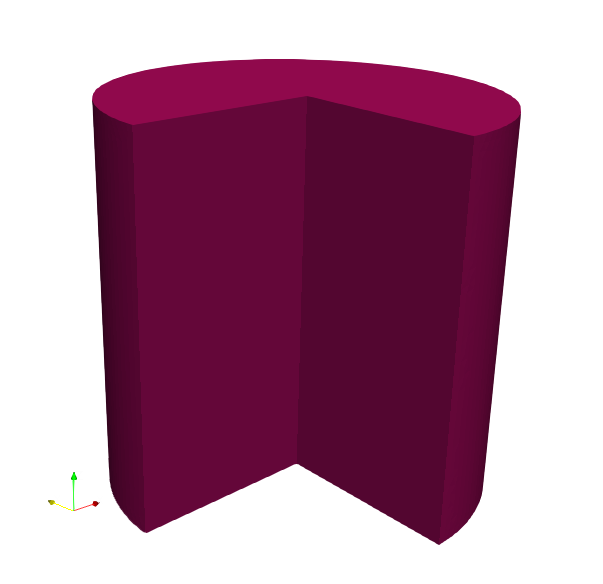}
	\includegraphics[width=0.45\textwidth]{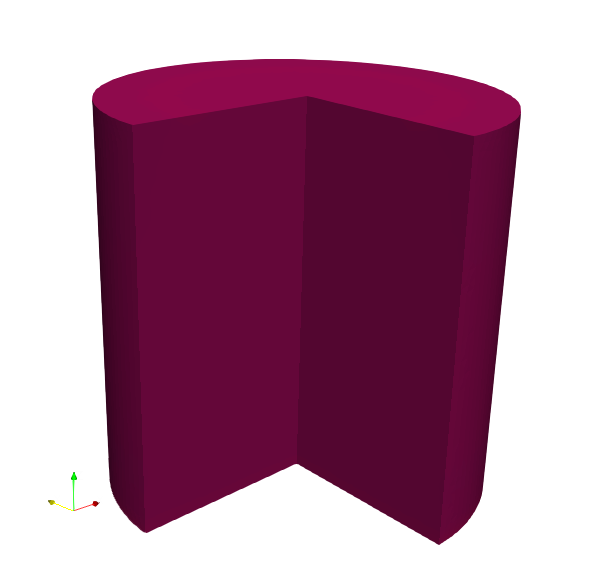}\\
	(a) $t = 0.0$ s $\hspace{3.75 cm}$ (b) $t = 0.2$ s\\
	\includegraphics[width=0.45\textwidth]{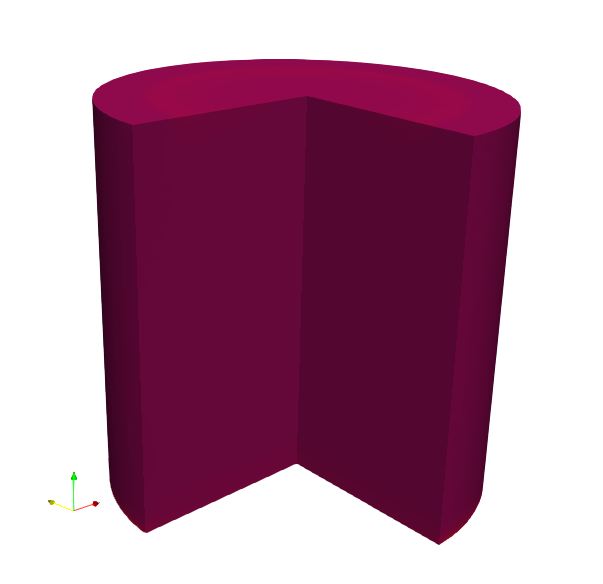}
	\includegraphics[width=0.45\textwidth]{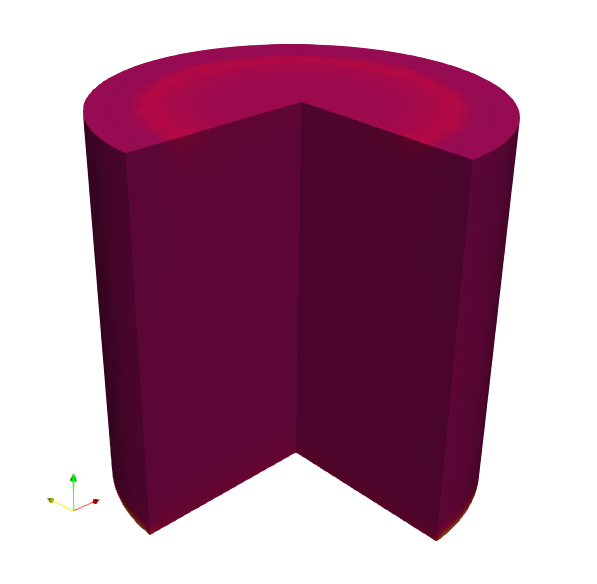}\\
	(c) $t = 0.4$ s $\hspace{3.75 cm}$ (d) $t = 0.6$ s\\
	\includegraphics[width=0.45\textwidth]{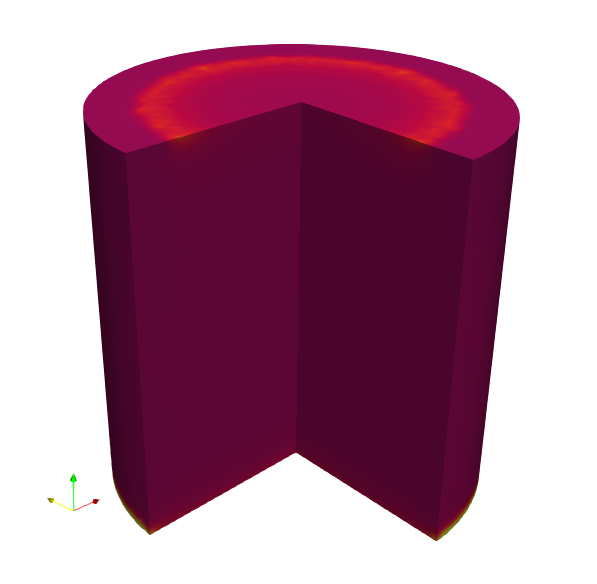}
	\includegraphics[width=0.45\textwidth]{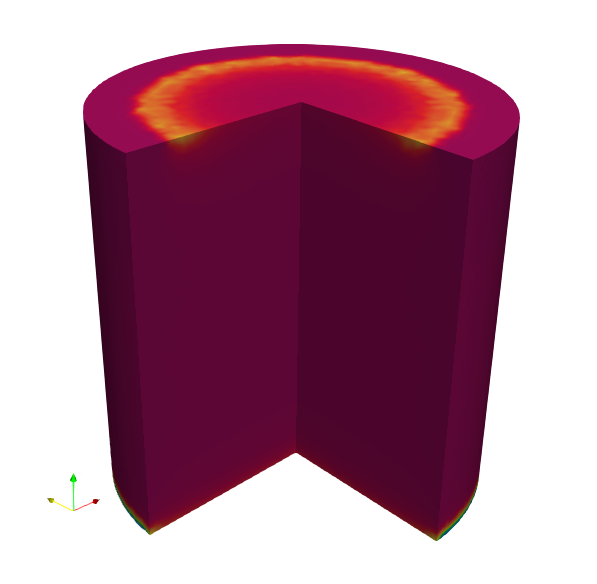}\\
	(e) $t = 0.8$ s $\hspace{3.75 cm}$ (f) $t = 1.0$ s\\
	\includegraphics[width=0.8\textwidth]{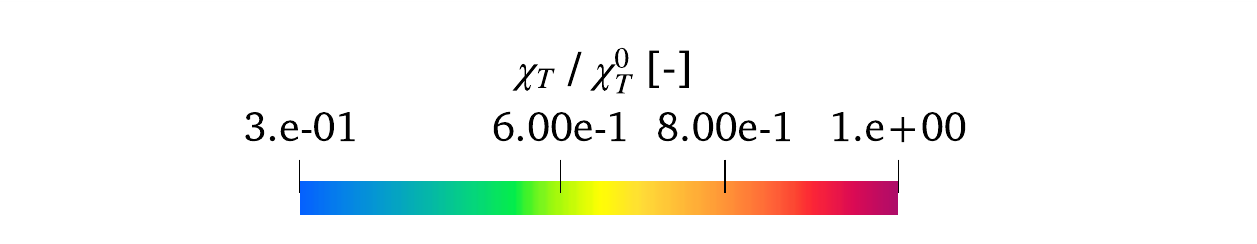}
	\caption{Example~\ref{example-elastic}: Normalized trapped hydrogen concentration at different time instants.}
	\label{fig_chiT_elastic}
\end{figure}

\begin{figure}
	\centering
	\includegraphics[width=0.45\textwidth]{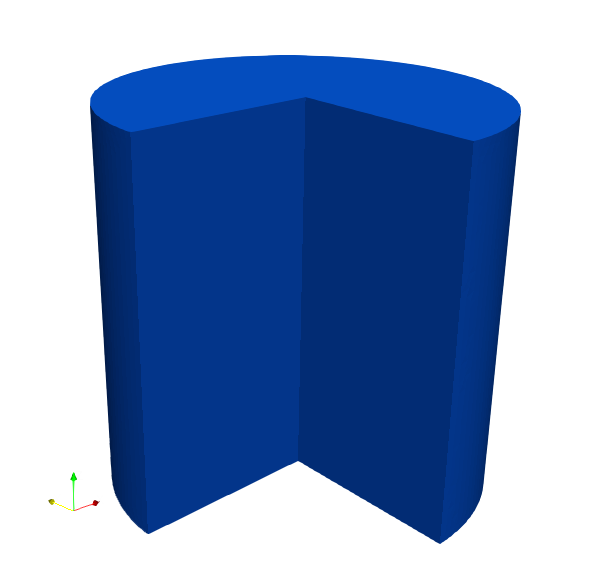}
	\includegraphics[width=0.45\textwidth]{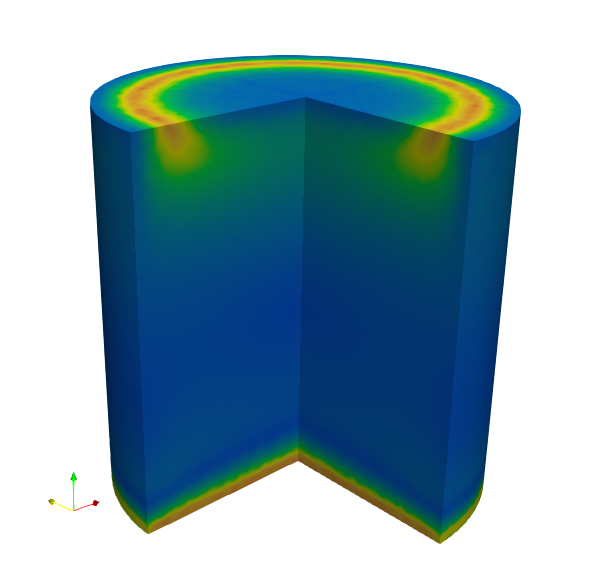}\\
	(a) $t = 0.0$ s $\hspace{3.75 cm}$ (b) $t = 0.2$ s\\
	\includegraphics[width=0.45\textwidth]{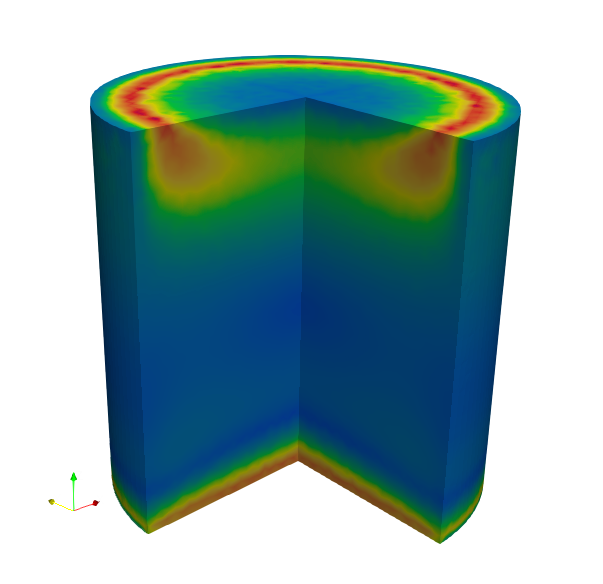}
	\includegraphics[width=0.45\textwidth]{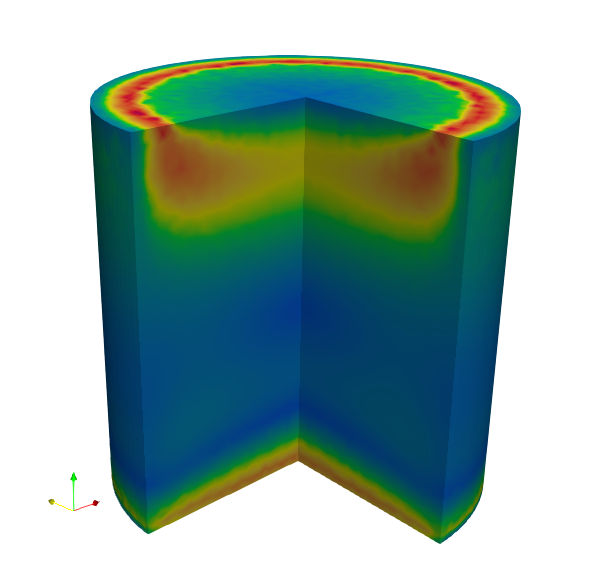}\\
	(c) $t = 0.4$ s $\hspace{3.75 cm}$ (d) $t = 0.6$ s\\
	\includegraphics[width=0.45\textwidth]{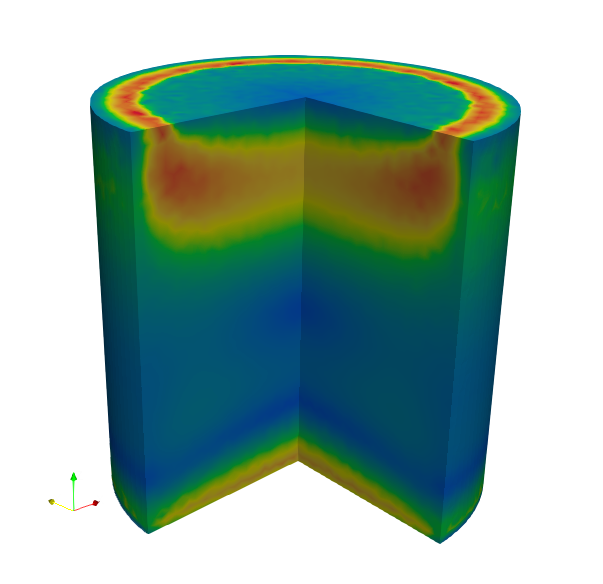}
	\includegraphics[width=0.45\textwidth]{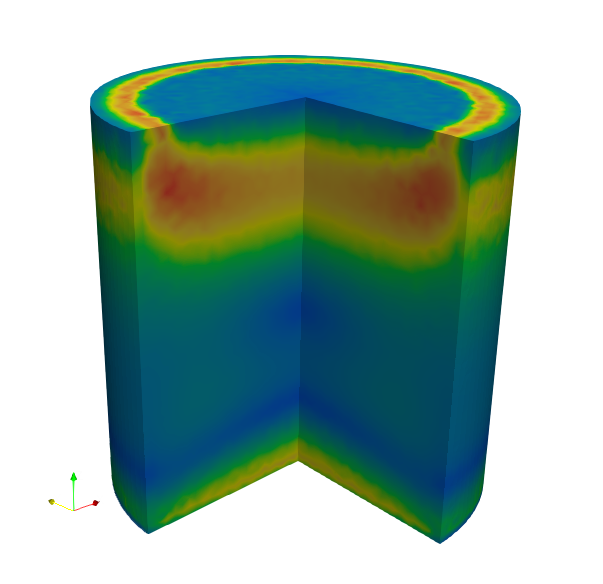}\\
	(e) $t = 0.8$ s $\hspace{3.75 cm}$ (f) $t = 1.0$ s\\
	\includegraphics[width=0.8\textwidth]{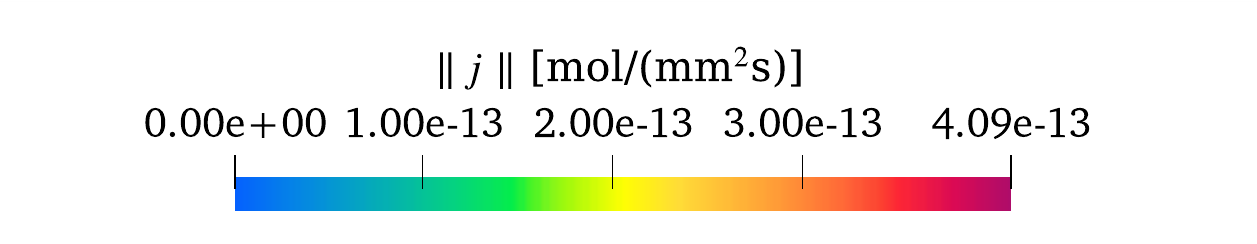}
	\caption{Example~\ref{example-elastic}: Norm of lattice mass flux at different time instants.}
	\label{fig_j_norm_elastic}
\end{figure}

\begin{figure}
	\centering
	\includegraphics[width=0.49\textwidth]{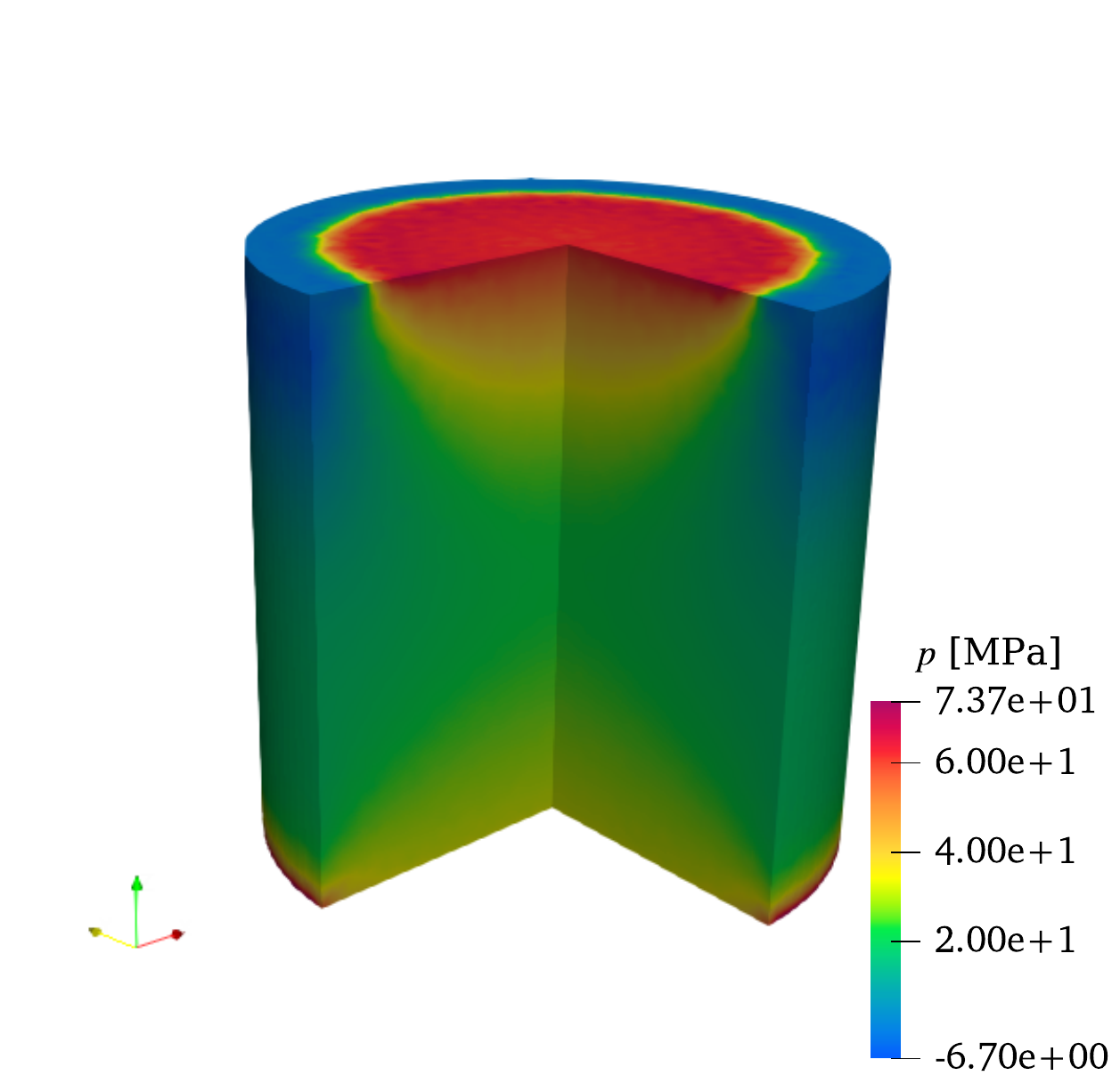}
	\includegraphics[width=0.49\textwidth]{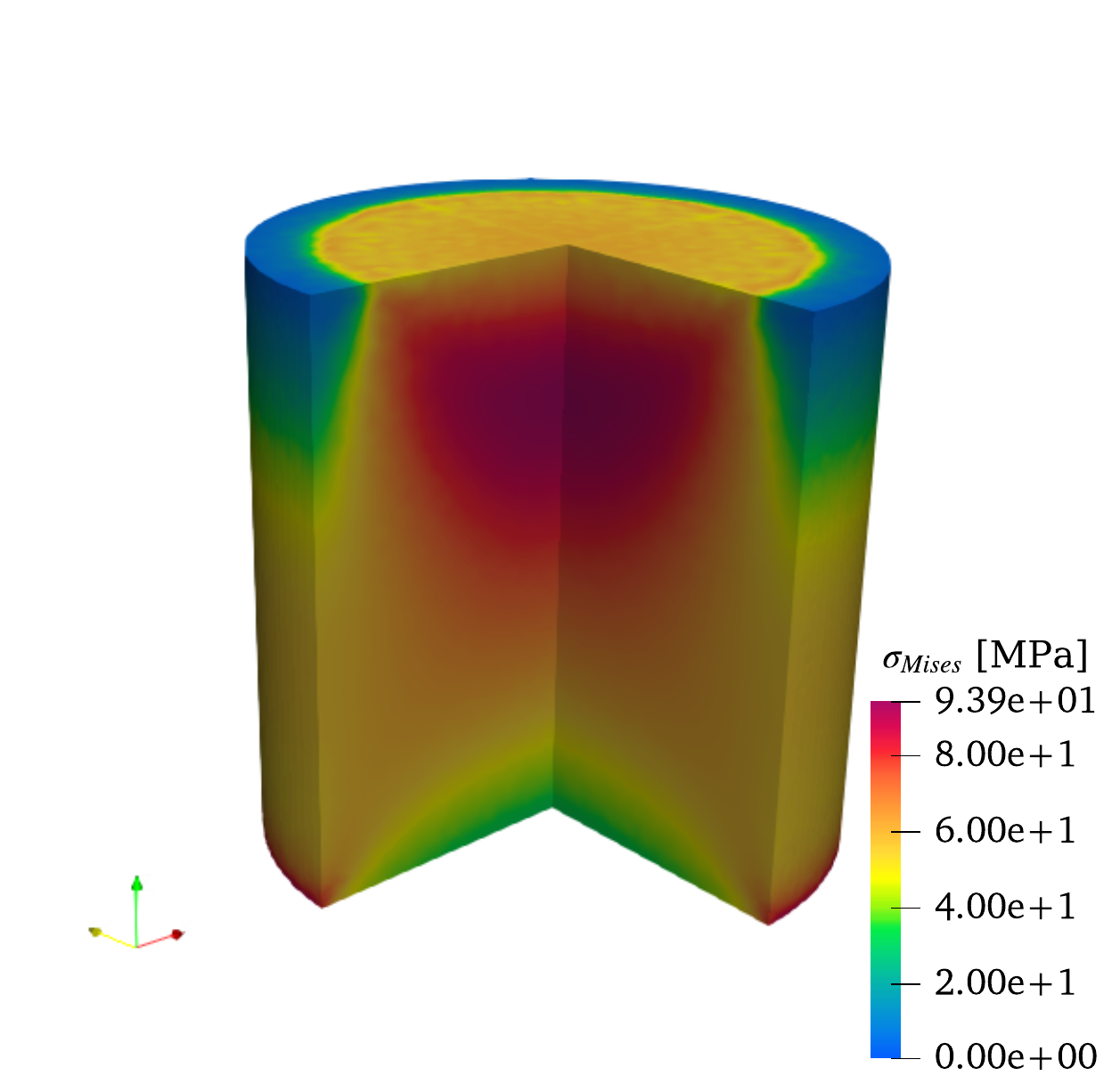}\\
	(a) Pressure [MPa] $\hspace{2.75 cm}$ (b) Von Mises stress [MPa]\\
	\includegraphics[width=0.49\textwidth]{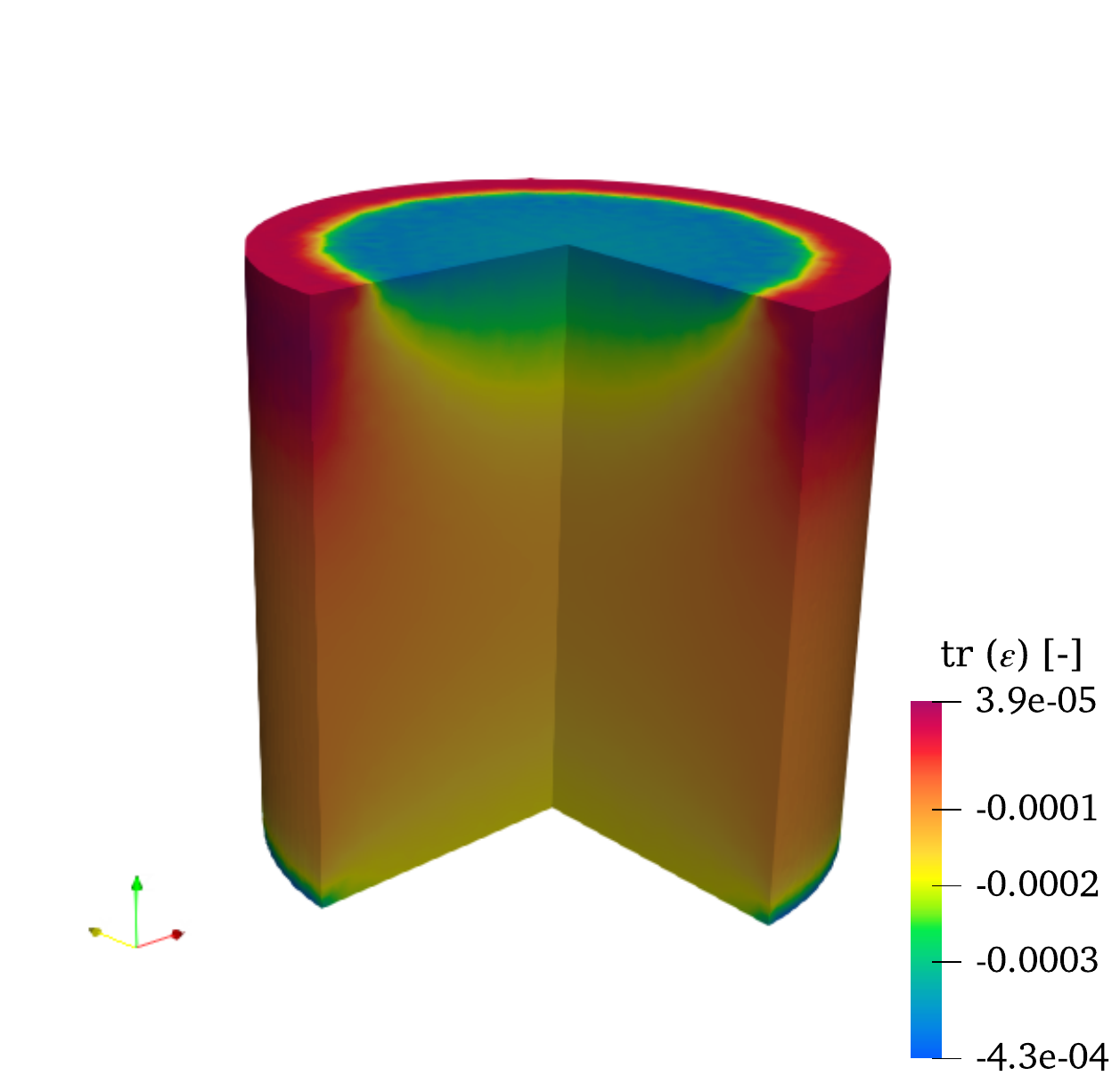}
	\includegraphics[width=0.49\textwidth]{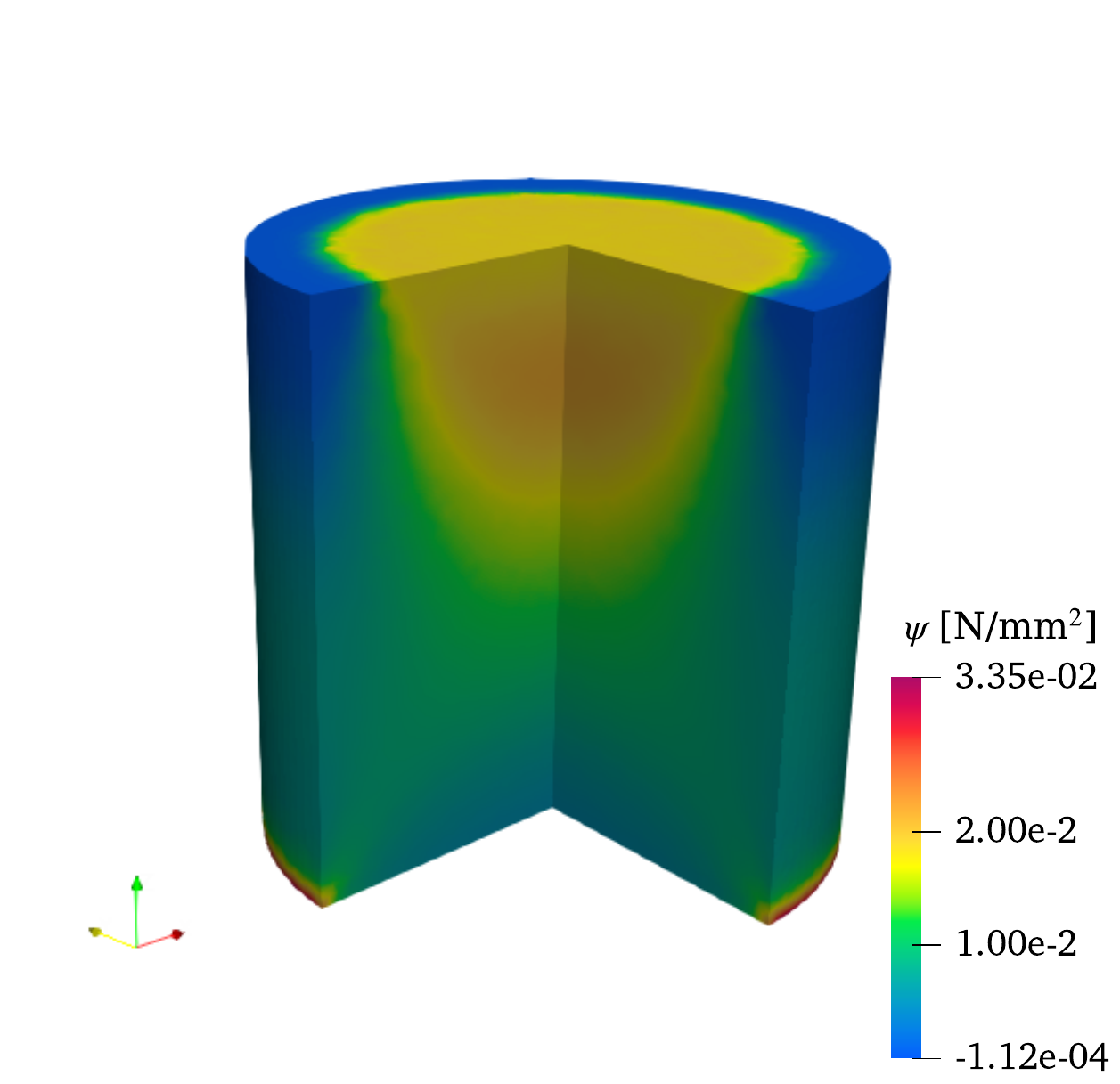}\\
	(c) Volumetric strain [ $-$ ] $\hspace{1.0 cm}$ (d) Helmholtz free energy [N/mm$^2$]\\
	\caption{Example~\ref{example-elastic}: Pressure (a), von Mises stress (b), volumetric strain (c), and Helmholtz free energy (d) at time $t=1$~s.}
	\label{fig_mech_elastic}
\end{figure}

Figure~\ref{fig_mu_elastic} illustrates the evolution of the chemical potential in the cylinder during the application of pressure. In turn, Figures~\ref{fig_chiL_elastic} and~\ref{fig_chiT_elastic} show, respectively, the concentrations of lattice and trapped hydrogen, normalized with $\chi_L^0$ and $\chi_T^0$, respectively.  In Figure~\ref{fig_j_norm_elastic} the norm of lattice hydrogen mass flux is depicted for several time instants. Figure~\ref{fig_mech_elastic} depicts the pressure (a), the von Mises stress (b), the volumetric strain (c), and the energy $\psi$ fields at time $t = 1$~s. All the solutions have been obtained with the variational method.

\begin{figure}
  \centering
  \includegraphics[width=0.9\textwidth]{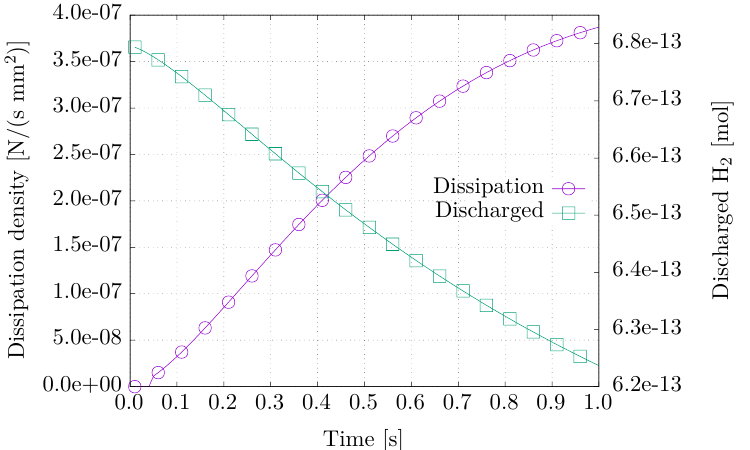}
  \caption{Example~\ref{example-elastic}: Dissipation density and NILS moles per unit volume vs. time.}
\label{fig_diss_mass_elastic}
\end{figure}

Figure~\ref{fig_diss_mass_elastic} plots the evolution of the dissipation density and the lattice hydrogen moles migrated due to the pressure application with the time. Coherently with the test performed, the NILS outflow decreases with time. Regarding the dissipation, although in an elastic evolution the dissipation is zero, we have an increasing and positive contribution of the diffusive phenomenon. We emphasize that the dissipation is non-negative for all time steps.

\begin{figure}
  \centering
  \includegraphics[width=0.9\textwidth]{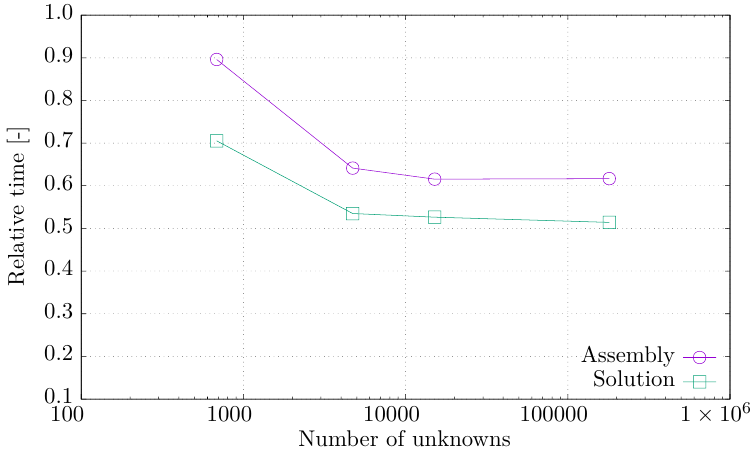}
  \caption{Example~\ref{example-elastic}: Assembly and solution times of variational method relative to the standard formulation as a function of the mesh size. }
\label{fig_times_elastic}
\end{figure}

Finally, Figure~\ref{fig_times_elastic} illustrates the CPU time spent in the tangent assembly and in the solution of the variational formulation relative to the standard one. It is confirmed that the assembly time is faster in the variational method. This is due to the fact that it is only necessary to compute and assemble approximately half of the entries in the tangent matrix. As far as the solution time is concerned, the results reveal that the unsymmetric solution is more expensive than the symmetric solution resulting from the variational method. In the largest problem tested, involving almost half a million unknowns, the average computational cost of the symmetric solver is close to half the cost of the unsymmetric one, as predicted by theoretical estimates.

\subsection{Plasticity coupled with hydrogen diffusion}
\label{example-plasticity}
In the second example we study hydrogen diffusion within a small strain, elastoplastic solid. We simulate a cylindrical volume of radius 0.5~mm and height 1~mm with the parameters  provided in Table~\ref{table_mat_parameters_elastic} extended to the plastic regime with the constants given in Table~\ref{table_mat_parameters_elastoplastic}.
\begin{table}[h]
	\caption{Material parameters for example~\ref{example-plasticity}.}
	\centering
    \renewcommand{\arraystretch}{1.2}
	\begin{tabular}{l c c c c}
	\toprule
	Parameter & Symbol & Value & Units\\
	\hline
    Yield stress & $\sigma_0$ & $250.0$ & MPa\\
	Isotropic hardening modulus & $H_{iso}$ & $60.0$ & MPa\\
	Kinematic hardening modulus & $H_{kin}$ & $60.0$ & MPa\\
	\bottomrule
	\end{tabular}
	\label{table_mat_parameters_elastoplastic}
\end{table}
The solid is subject to a uniform pressure ramped up to a maximum value of~$600$~MPa at a constant rate during~1~second on the central circular surface of radius~0.02~mm located on the upper surface, while precluding any displacement in the bottom surface and any change in the chemical potential in the bottom and the curved surfaces.

For the numerical model, only one fourth of cylinder is modeled, and 19040 hexahedral elements are employed. To accommodate properly the compressibility constraint of the plastic flow, we use an assumed strain, mean dilatation formulation. On the symmetry surface of the $xz$ plane, the displacement in the $y$-direction is set to zero. Also, the displacement in the $x$-direction of the symmetry surface of the $yz$ plane is constrained. We obtain the solutions employing the variational formulation introduced in this work with a constant time step size equal to 0.005~s.

\begin{figure}
	\centering
\includegraphics[width=0.45\textwidth]{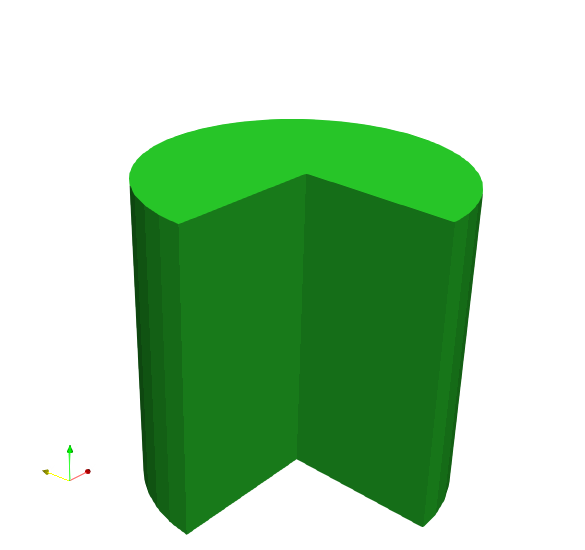}
	\includegraphics[width=0.45\textwidth]{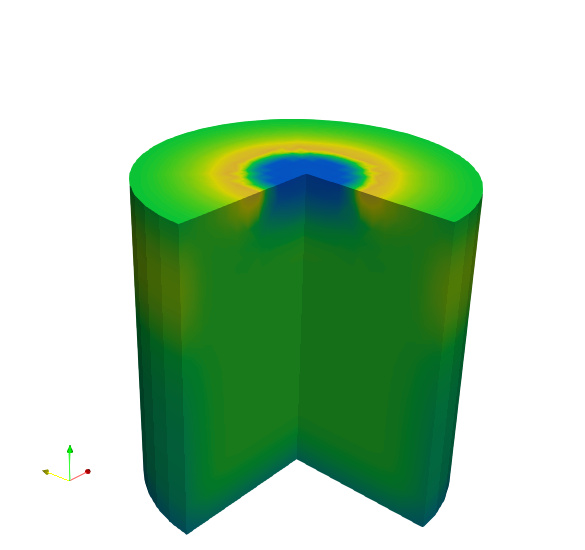}\\
	(a) $t = 0.0$ s $\hspace{3.75 cm}$ (b) $t = 0.2$ s\\
	\includegraphics[width=0.49\textwidth]{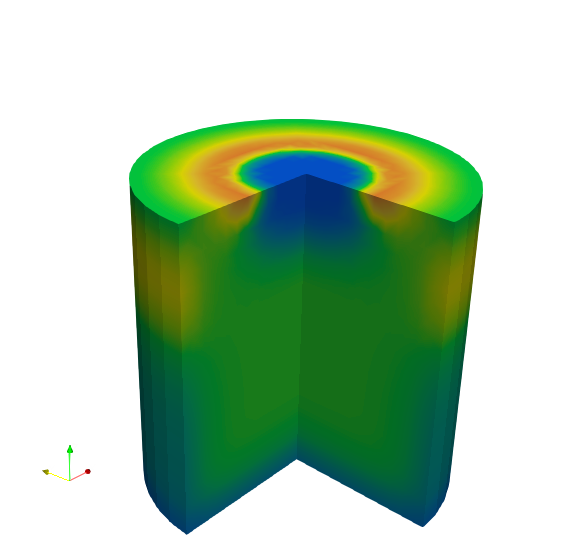}
	\includegraphics[width=0.45\textwidth]{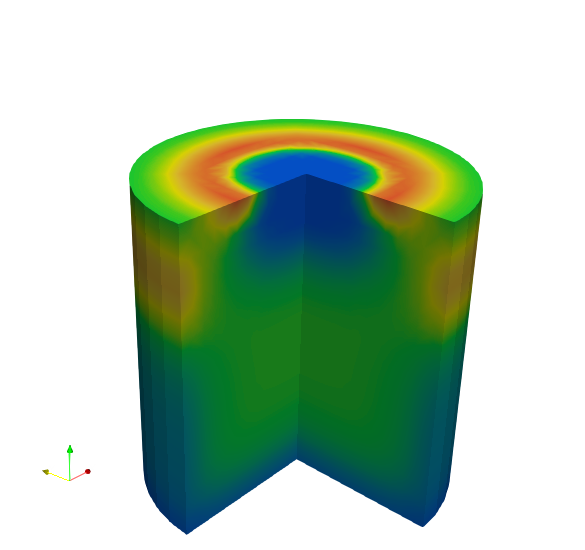}\\
	(c) $t = 0.4$ s $\hspace{3.75 cm}$ (d) $t = 0.6$ s\\
	\includegraphics[width=0.45\textwidth]{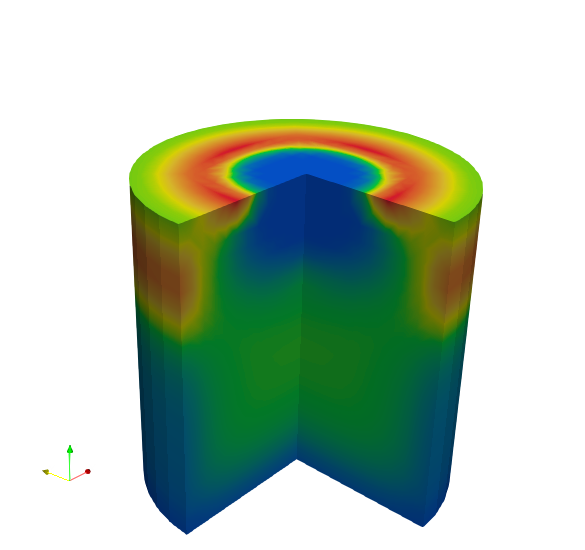}
	\includegraphics[width=0.45\textwidth]{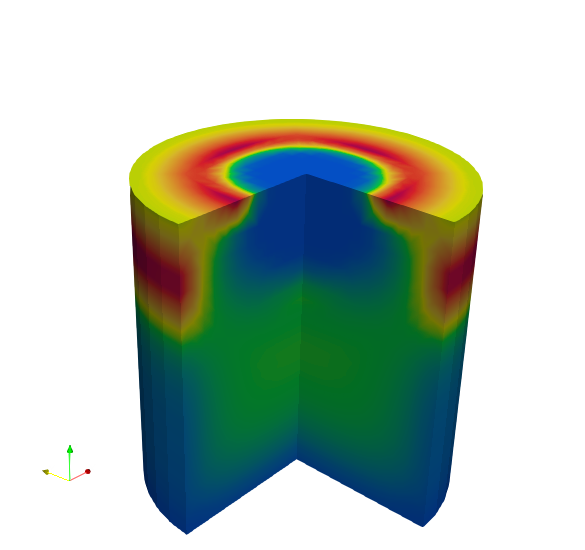}\\
	(e) $t = 0.8$ s $\hspace{3.75 cm}$ (f) $t = 1.0$ s\\
	\includegraphics[width=0.8\textwidth]{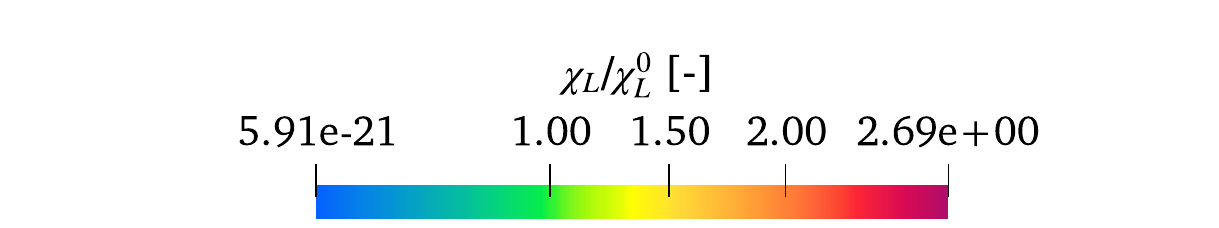}
	\caption{Example~\ref{example-plasticity}: Normalized lattice hydrogen concentration at different time instants.}
	\label{fig_chiL_elastoplastic}
\end{figure}

\begin{figure}
	\centering
	\includegraphics[width=0.45\textwidth]{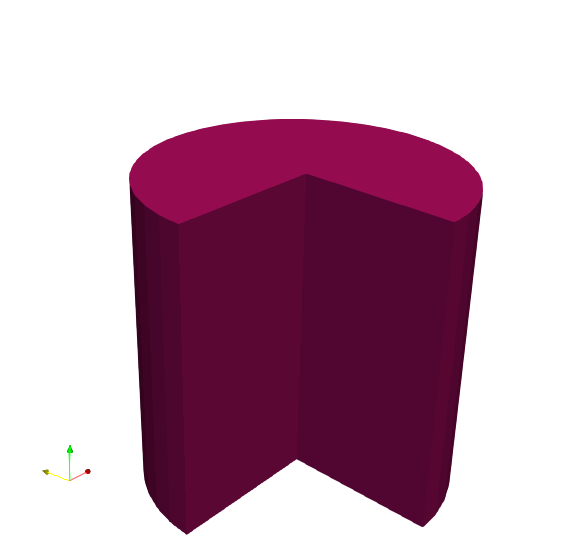}
	\includegraphics[width=0.45\textwidth]{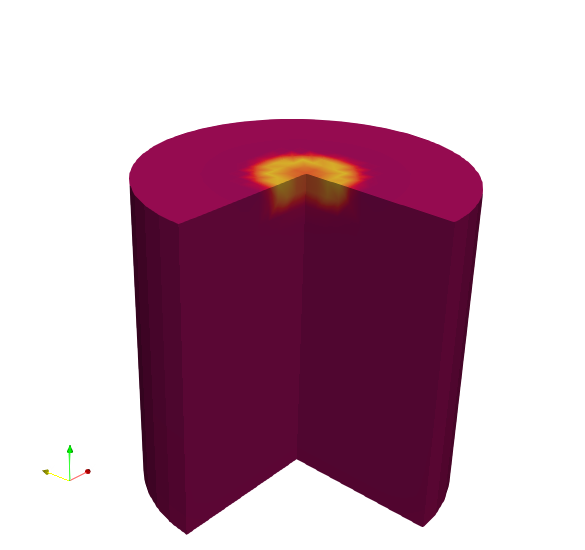}\\
	(a) $t = 0.0$ s $\hspace{3.75 cm}$ (b) $t = 0.2$ s\\
	\includegraphics[width=0.45\textwidth]{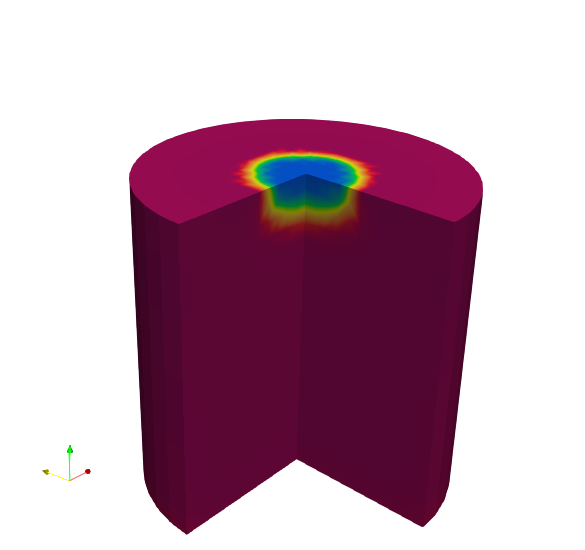}
	\includegraphics[width=0.45\textwidth]{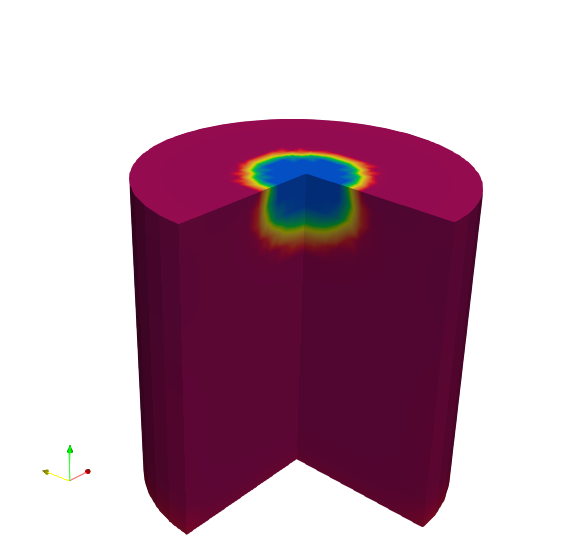}\\
	(c) $t = 0.4$ s $\hspace{3.75 cm}$ (d) $t = 0.6$ s\\
	\includegraphics[width=0.45\textwidth]{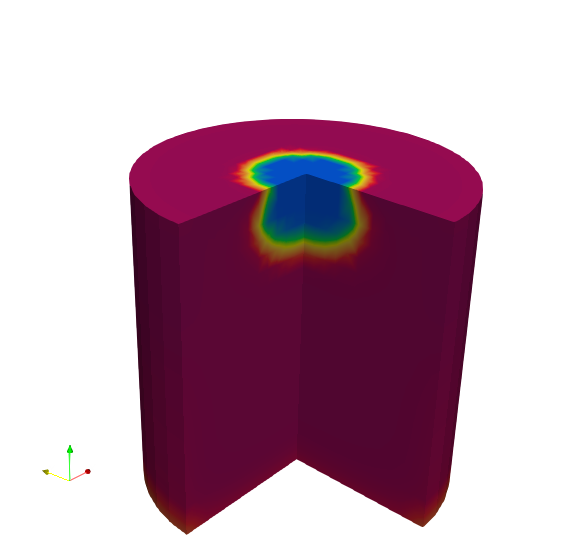}
	\includegraphics[width=0.45\textwidth]{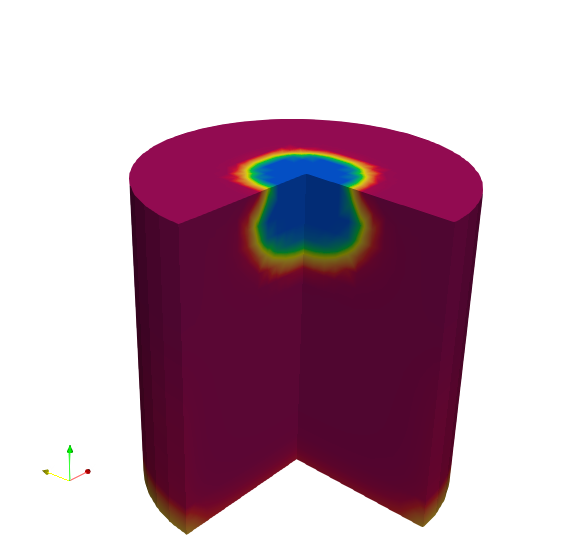}\\
	(e) $t = 0.8$ s $\hspace{3.75 cm}$ (f) $t = 1.0$ s\\
	\includegraphics[width=0.8\textwidth]{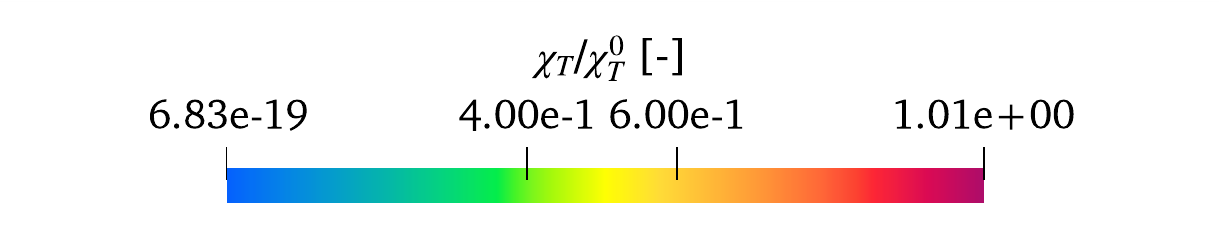}
	\caption{Example~\ref{example-plasticity}: Normalized trapped hydrogen concentration at different time instants.}
	\label{fig_chiT_elastoplastic}
\end{figure}

\begin{figure}
	\centering
	\includegraphics[width=0.45\textwidth]{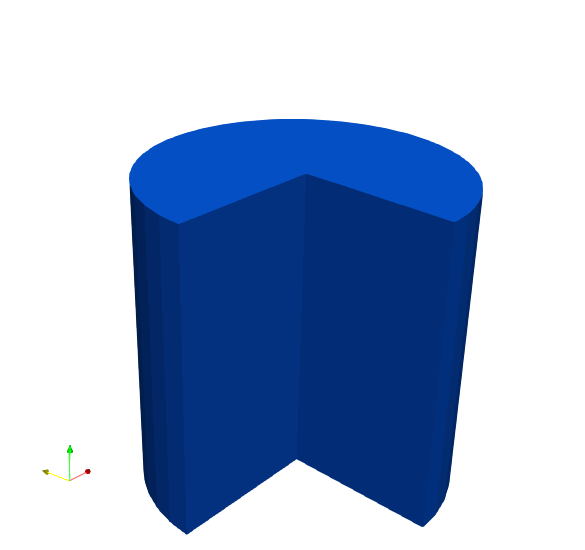}
	\includegraphics[width=0.45\textwidth]{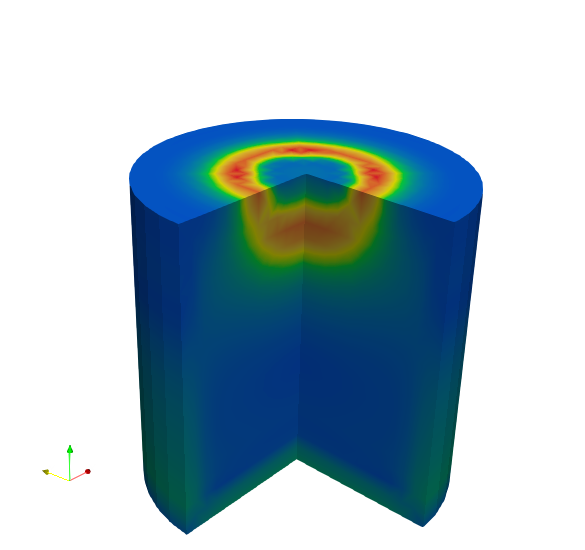}\\
	(a) $t = 0.0$ s $\hspace{3.75 cm}$ (b) $t = 0.2$ s\\
	\includegraphics[width=0.45\textwidth]{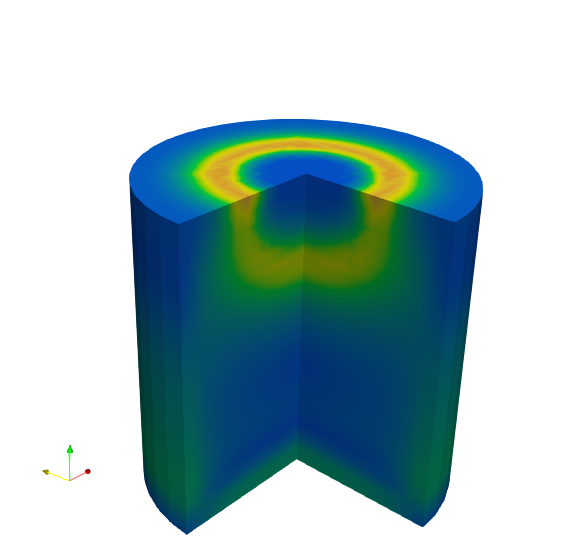}
	\includegraphics[width=0.45\textwidth]{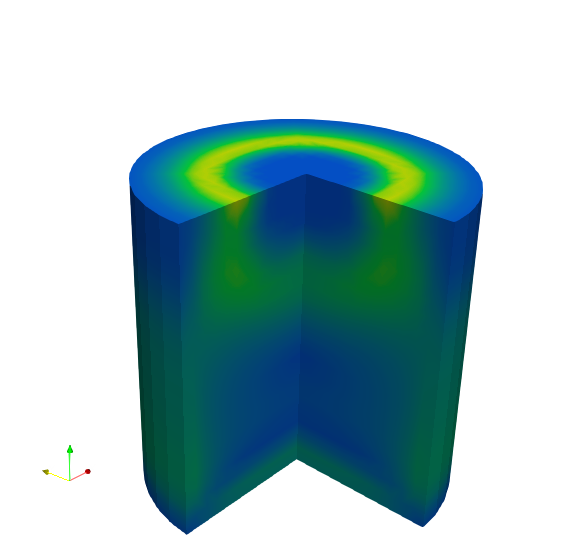}\\
	(c) $t = 0.4$ s $\hspace{3.75 cm}$ (d) $t = 0.6$ s\\
	\includegraphics[width=0.45\textwidth]{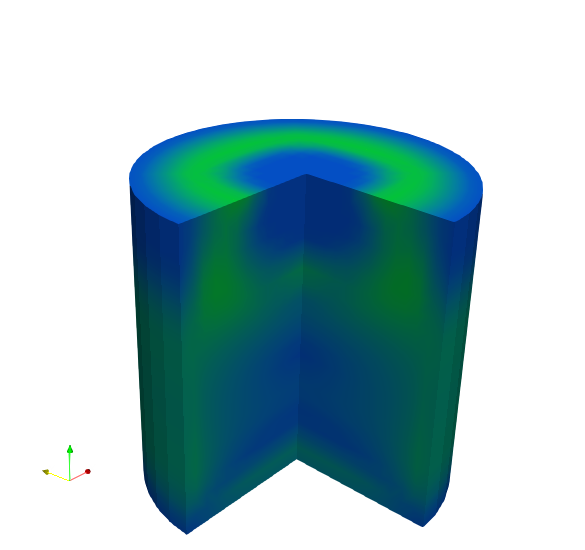}
	\includegraphics[width=0.45\textwidth]{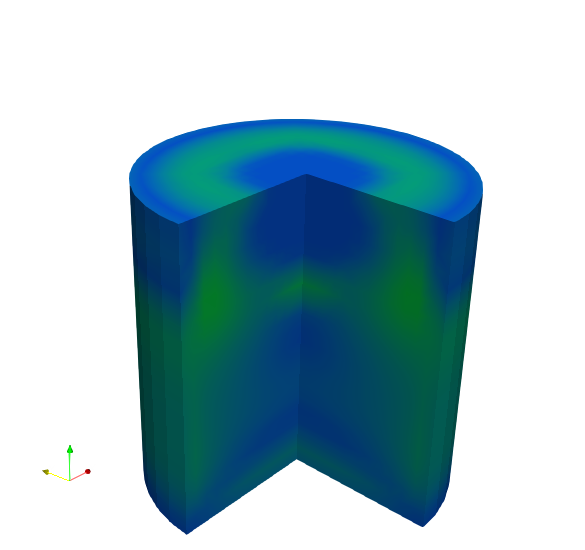}\\
	(e) $t = 0.8$ s $\hspace{3.75 cm}$ (f) $t = 1.0$ s\\
	\includegraphics[width=0.8\textwidth]{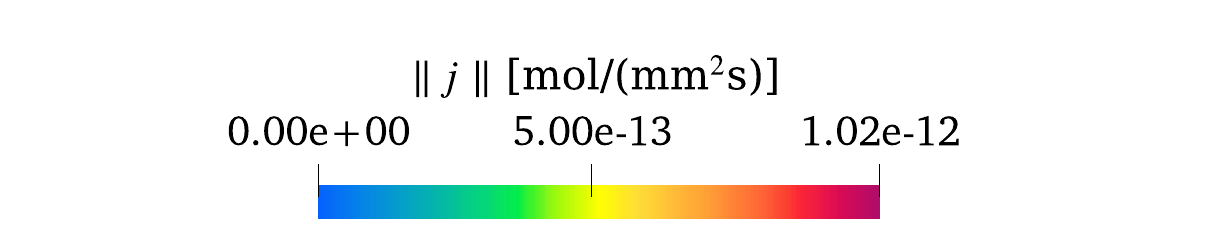}
	\caption{Example~\ref{example-plasticity}: Norm of mass flux at different time instants.}
	\label{fig_mass_flux_elastoplastic}
\end{figure}

\begin{figure}
	\centering
	\includegraphics[width=0.40\textwidth]{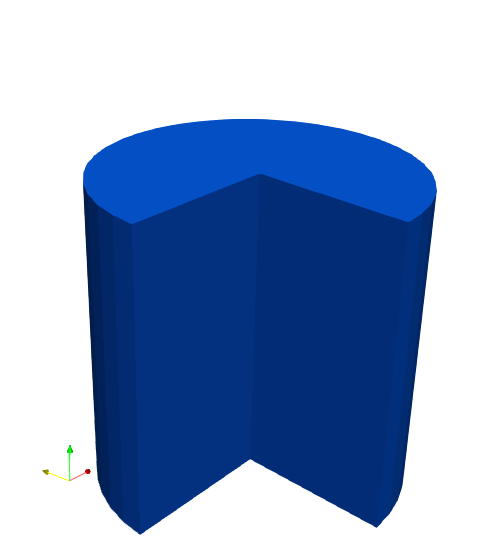}
	\includegraphics[width=0.40\textwidth]{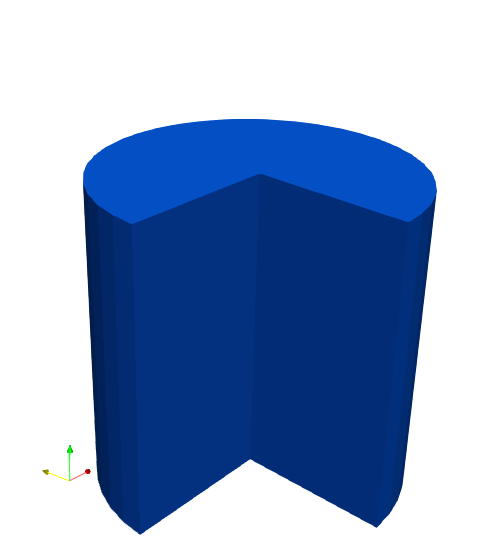}\\
	(a) $t = 0.0$ s $\hspace{3.75 cm}$ (b) $t = 0.2$ s\\
	\includegraphics[width=0.40\textwidth]{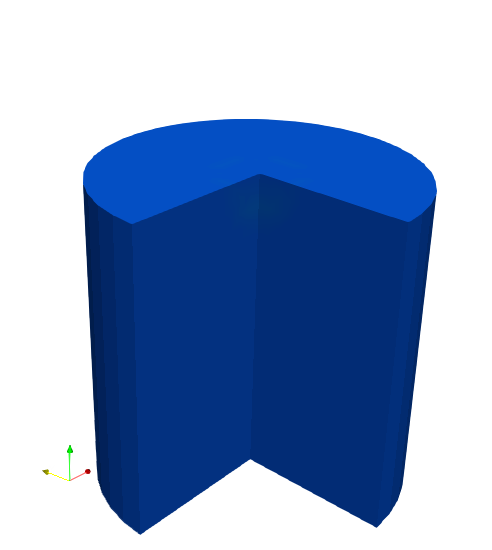}
	\includegraphics[width=0.40\textwidth]{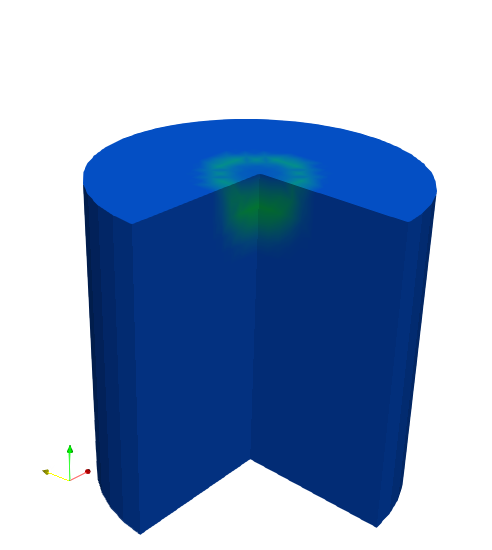}\\
	(c) $t = 0.4$ s $\hspace{3.75 cm}$ (d) $t = 0.6$ s\\
	\includegraphics[width=0.40\textwidth]{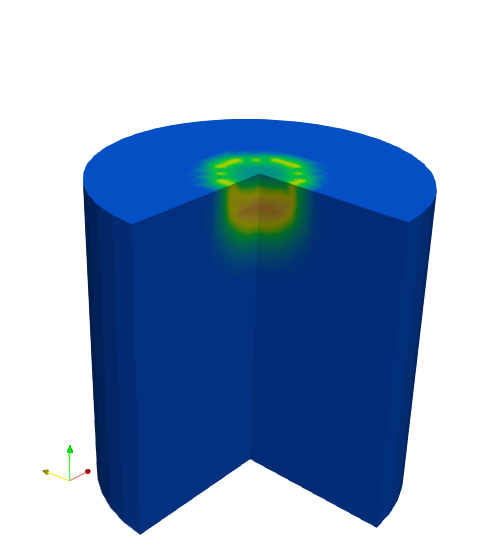}
	\includegraphics[width=0.40\textwidth]{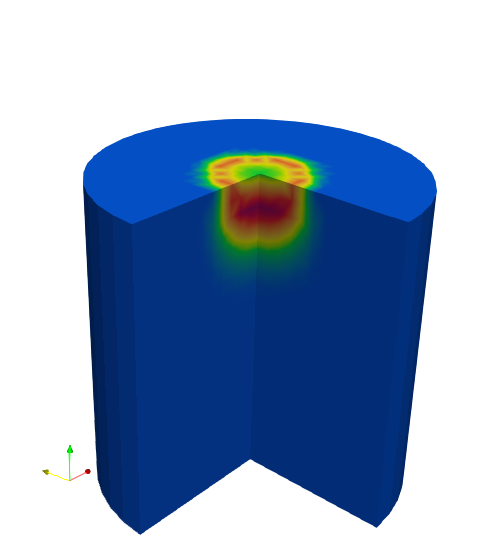}\\
	(e) $t = 0.8$ s $\hspace{3.75 cm}$ (f) $t = 1.0$ s\\
	\includegraphics[width=0.8\textwidth]{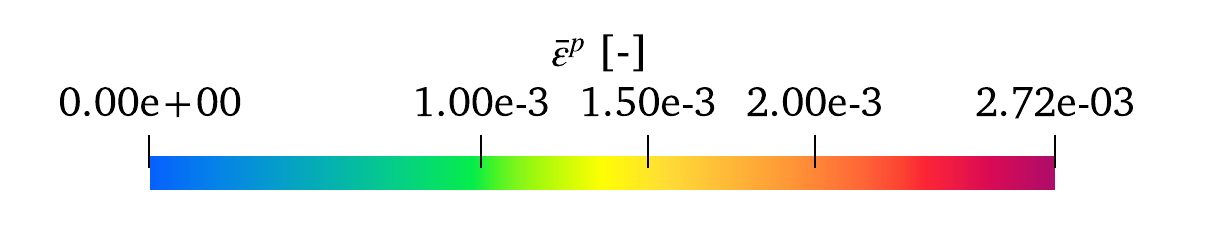}
	\caption{Example~\ref{example-plasticity}: Plastic slip at different time instants.}
	\label{fig_pslip_elastoplastic}
\end{figure}

\begin{figure}
	\centering
	\includegraphics[width=0.45\textwidth]{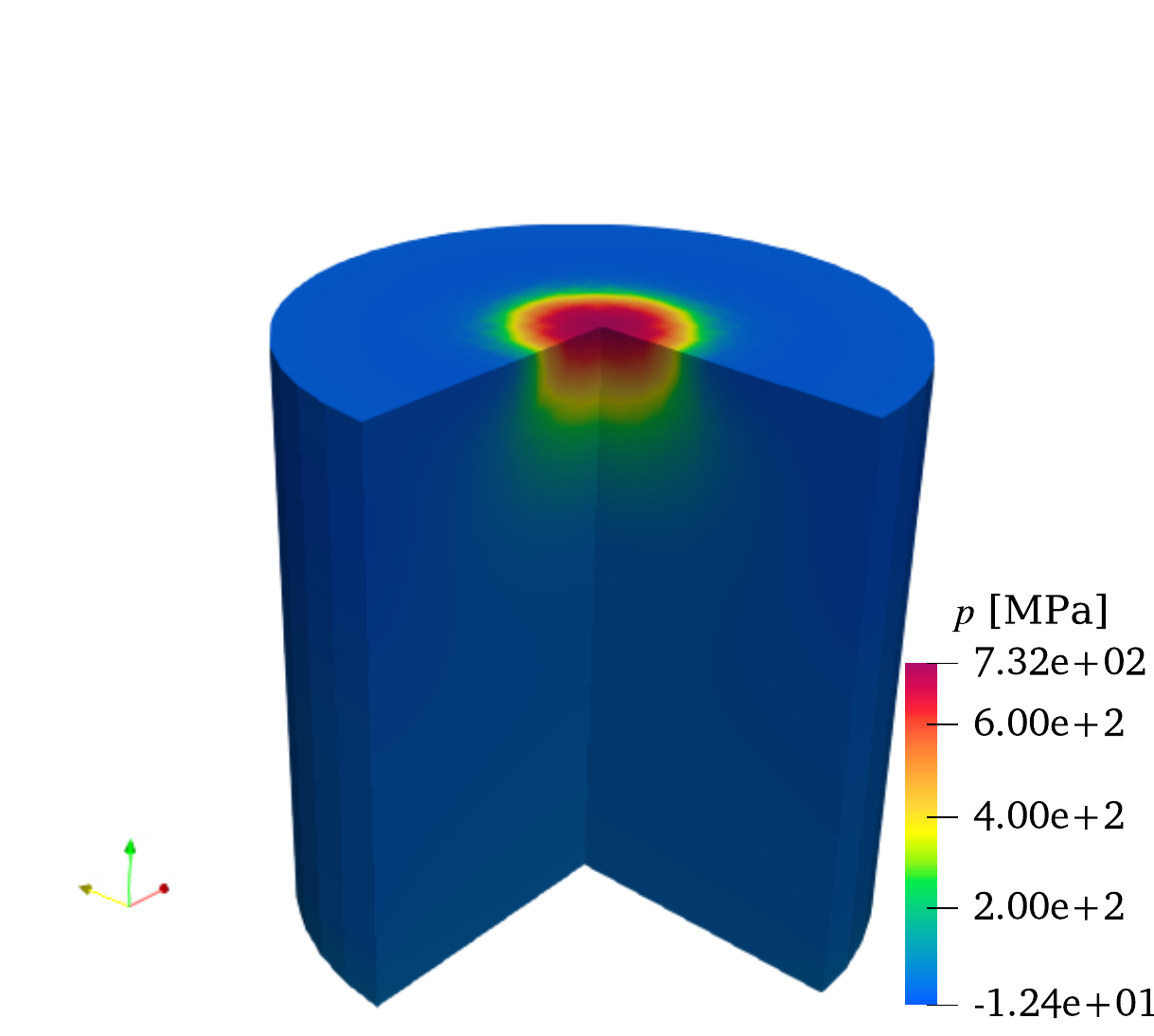}
	\includegraphics[width=0.45\textwidth]{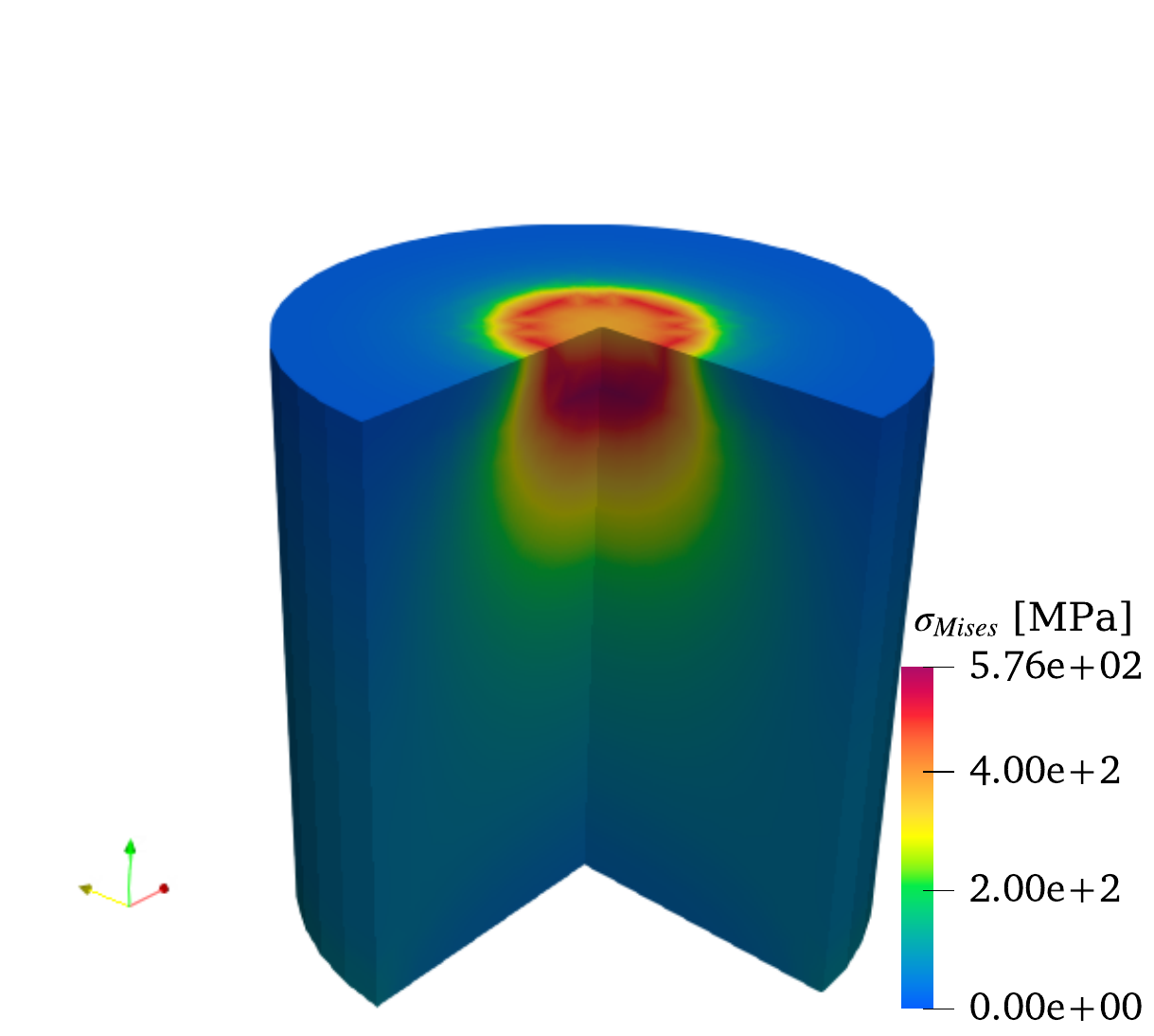}\\
	(a) Pressure [MPa] $\hspace{2.75 cm}$ (b) Von Mises stress [MPa]\\
	\includegraphics[width=0.45\textwidth]{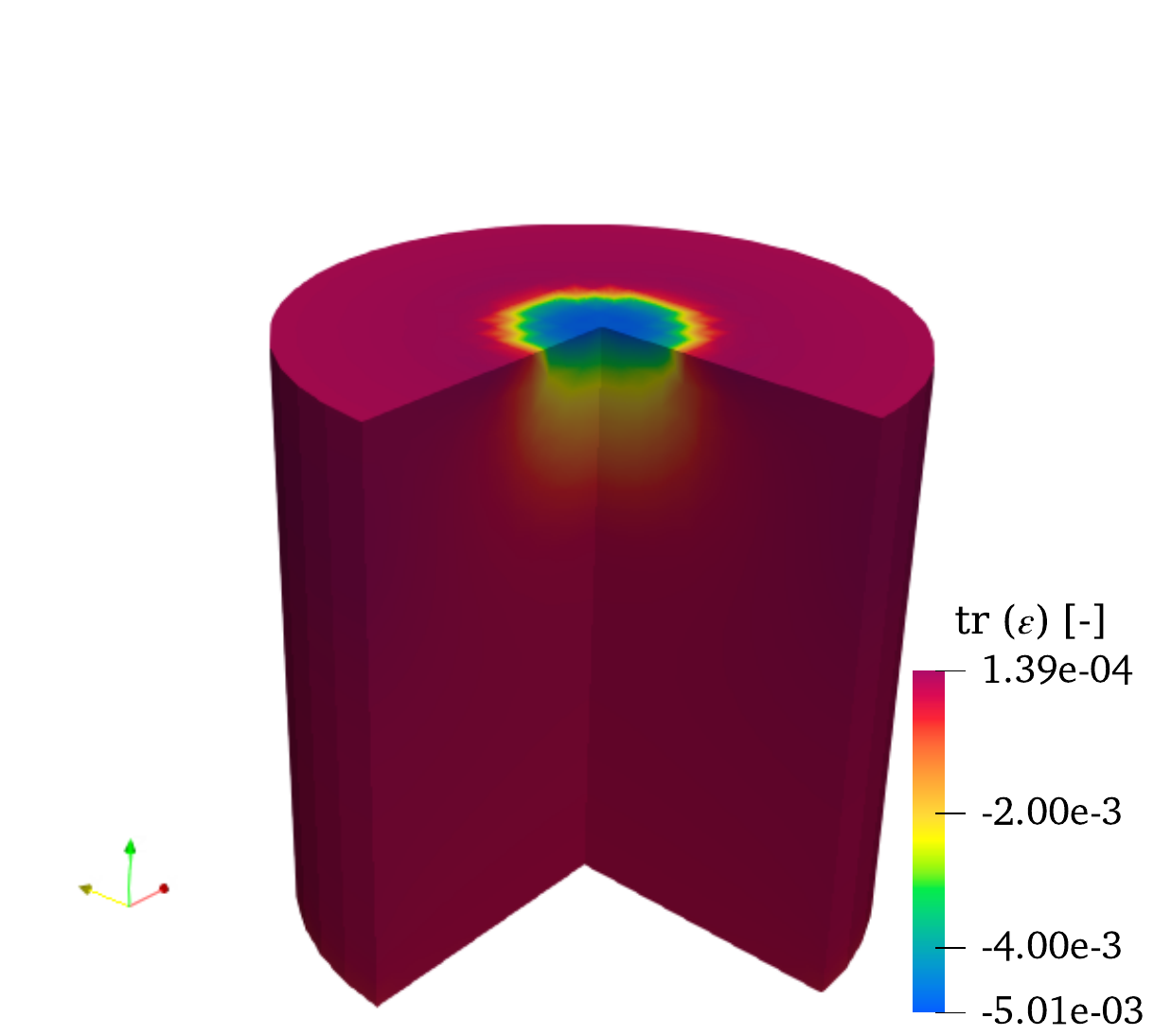}
	\includegraphics[width=0.45\textwidth]{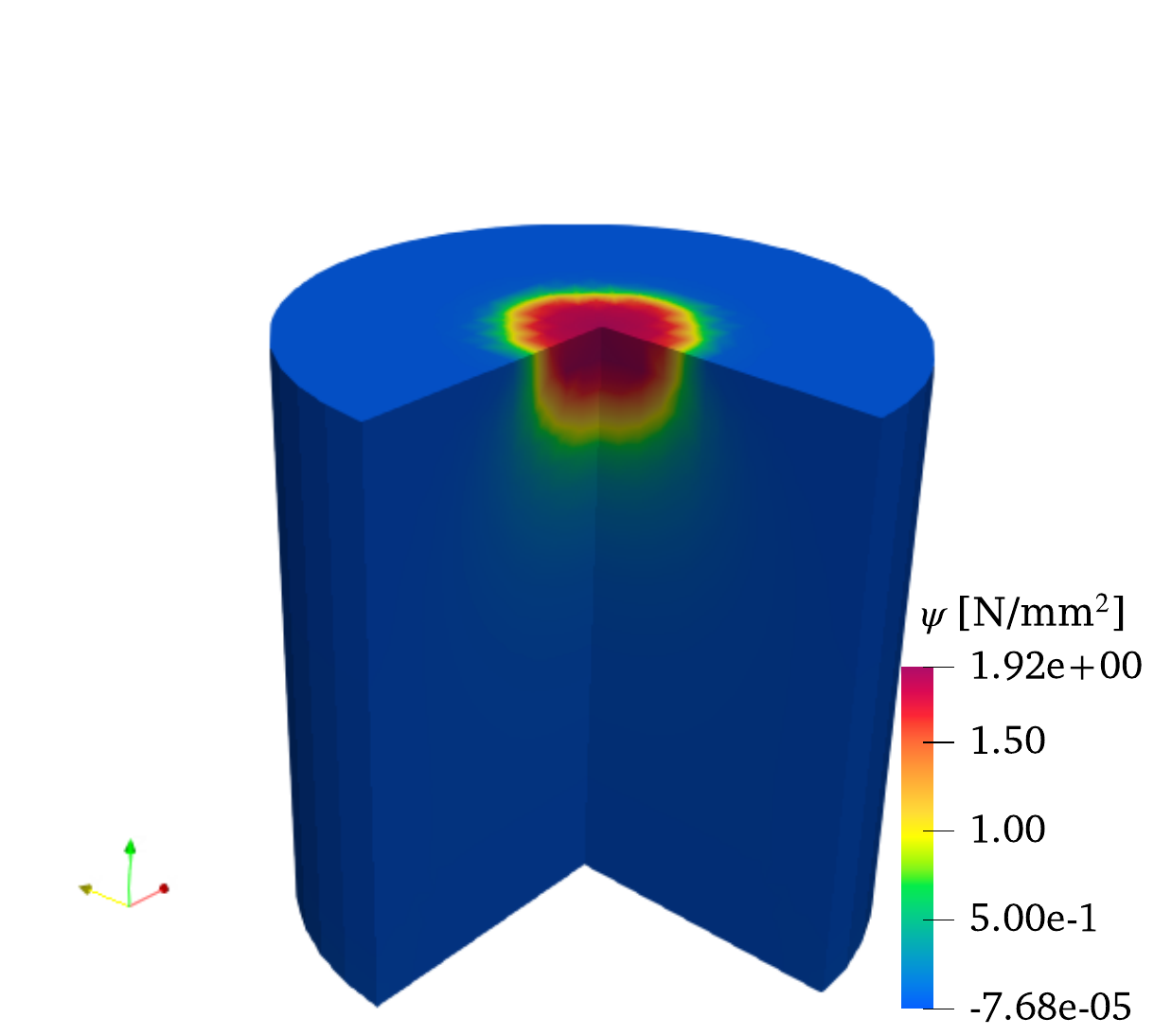}\\
	(c) Volumetric strain [ $-$ ] $\hspace{1.0 cm}$ (d) Helmholtz free energy [N/mm$^2$]\\
	\caption{Example~\ref{example-plasticity}: Pressure (a), von Mises stress (b), volumetric strain (c), and Helmholtz free energy (d) at time $t = 1$ s.}
	\label{fig_mech_elastoplastic}
\end{figure}

Figures~\ref{fig_chiL_elastoplastic} and~\ref{fig_chiT_elastoplastic} show, respectively, the normalized lattice and trapped hydrogen concentrations at different time steps. Figures~\ref{fig_mass_flux_elastoplastic} and~\ref{fig_pslip_elastoplastic} illustrate the evolution of the norm of the mass flux and the plastic slip, respectively, during the application of pressure. Comparing figures~\ref{fig_chiT_elastoplastic} and~\ref{fig_pslip_elastoplastic}, we confirm the relationship between the plastic deformation and the trapped hydrogen in the solid (see  Eqs.~\ref{nt_dependency_eq} and~\ref{nt_expression_eq}). Finally, the pressure (a), the von Mises stress (b), the volumetric strain (c), and the Helmholtz free energy (d) fields at time $t=1$~s are depicted in Figure~\ref{fig_mech_elastoplastic}.  

\begin{figure}
  \centering
  \includegraphics[width=0.9\textwidth]{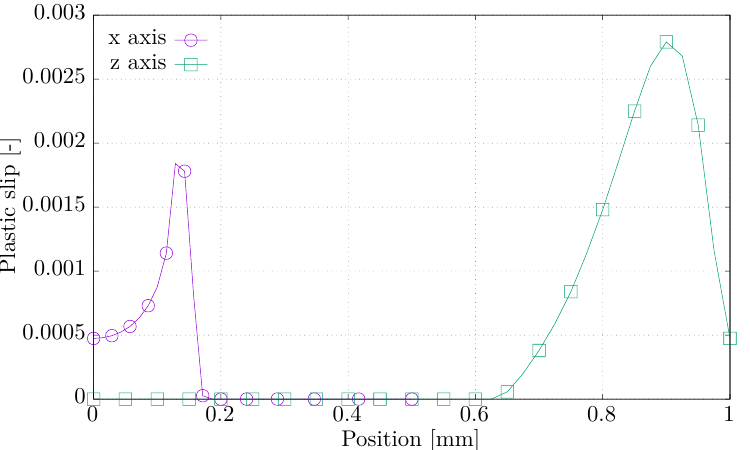}
  \caption{Example~\ref{example-plasticity}: Distribution of plastic slip through the radial direction in the horizontal plane.}
\label{fig_plastic_slip_elastoplastic}
\end{figure}

Figure~\ref{fig_plastic_slip_elastoplastic} shows the plastic slip distribution along the cylinder axis, $z$-axis, and along a radial direction parallel to $x$-axis contained in the top plane. In both plots, the maximum plastic slip coincides with the maximum stress, as expected (see Figures~\ref{fig_pslip_elastoplastic} and~\ref{fig_mech_elastoplastic} (b)). Note that the normalized concentrations, in particular the normalized trapped hydrogen concentration, exhibit their lowest values at these locations (see Figures~\ref{fig_chiL_elastoplastic} and \ref{fig_chiT_elastoplastic}), which is coherent with the solved problem. After those points, they evolve to satisfy the boundary conditions.

\begin{figure}
  \centering
  \includegraphics[width=0.9\textwidth]{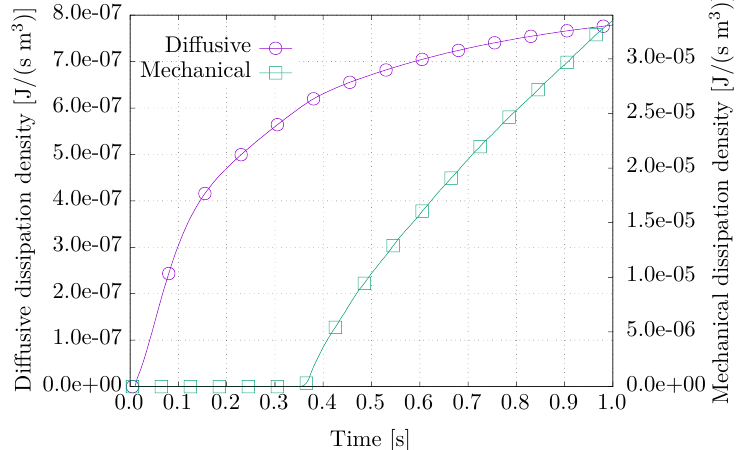}
  \caption{Example~\ref{example-plasticity}: Dissipation rate per unit volume: mechanical and diffusive contributions.}
\label{fig_diss_elastoplastic}
\end{figure}

Figure~\ref{fig_diss_elastoplastic} depicts the two contributions to the dissipation density: the mechanical term, due to plastic dissipation, and the contribution resulting from hydrogen diffusion. It is observed that 
the mechanical dissipation vanishes until the outset of plastic deformation, as expected. As far as the diffusive contribution is concerned, we note that its slope is higher in the first part of the simulation, until $t = 0.4$~s, coherently with the evolution of the mass flux, as it can be confirmed in Figure~\ref{fig_mass_flux_elastoplastic}. Both dissipation contributions are non-negative, as the theory predicts.

\subsection{Analysis of hydrogen transport in a notched specimen during a tensile test}
\label{example-notched-specimen}
To fully exploit the generality of our method, particularly in the plastic regime, the third example we propose is the numerical analysis of hydrogen transport in a notched specimen during a tensile test. In our simulation, we employ a parallelepiped specimen, notched in the symmetry $xz$ plane, with a height of 1~mm and  thickness of 0.025~mm. In its wider part, the specimen width is 0.30~mm. In the notched area, the width is 0.20~mm. We simulate only one-fourth of the specimen. On the symmetry $x$ and $y$ axes, the displacements in the $y$-direction and $x$-direction, respectively, are set to zero. Along the upper surface of the specimen, we prescribe a vertical uniform displacement, constantly increased up to a maximum value of $10^{-3}$~mm during 1~s. We address the case in which there is no supply of hydrogen through the outer boundary. Thus, the total hydrogen content in the domain is constant to its initial value. As stated in previous sections, our aim here is not to perform a thorough analysis of this mechanical problem but rather to illustrate the use of the chemical potential as the independent variable. Hence, we must prescribe the Neumann boundary conditions for this quantity, rather than imposing a value for the gradient of the lattice hydrogen concentration. We simulate such an insulation imposing $\mbs{j} \cdot \mbs{n} = 0$.

\begin{figure}
  \centering
  \includegraphics[width=0.9\textwidth]{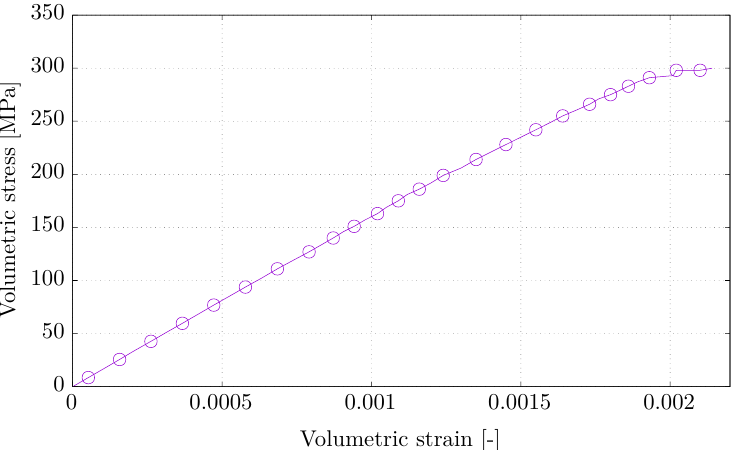}
	\caption{Example~\ref{example-notched-specimen}: Strain-stress curve.}
	\label{fig_tensile_test_notch}
\end{figure}

Our finite-element mesh consists of 9744 hexahedral elements, as in example \ref{example-plasticity}. The parameters used in this set of simulations are the same as in Example~\ref{example-plasticity}.
Figure~\ref{fig_tensile_test_notch} shows the strain-stress curve for this test. This work is not aimed to study the fracture, therefore the simulation is conducted only to reach the plastic regime. Our simulations reproduce the expected results for the material parameters, which indicates that our finite element implementation, with respect to the mechanical response, is correct.

\begin{figure}
	\centering
\includegraphics[width=0.45\textwidth]{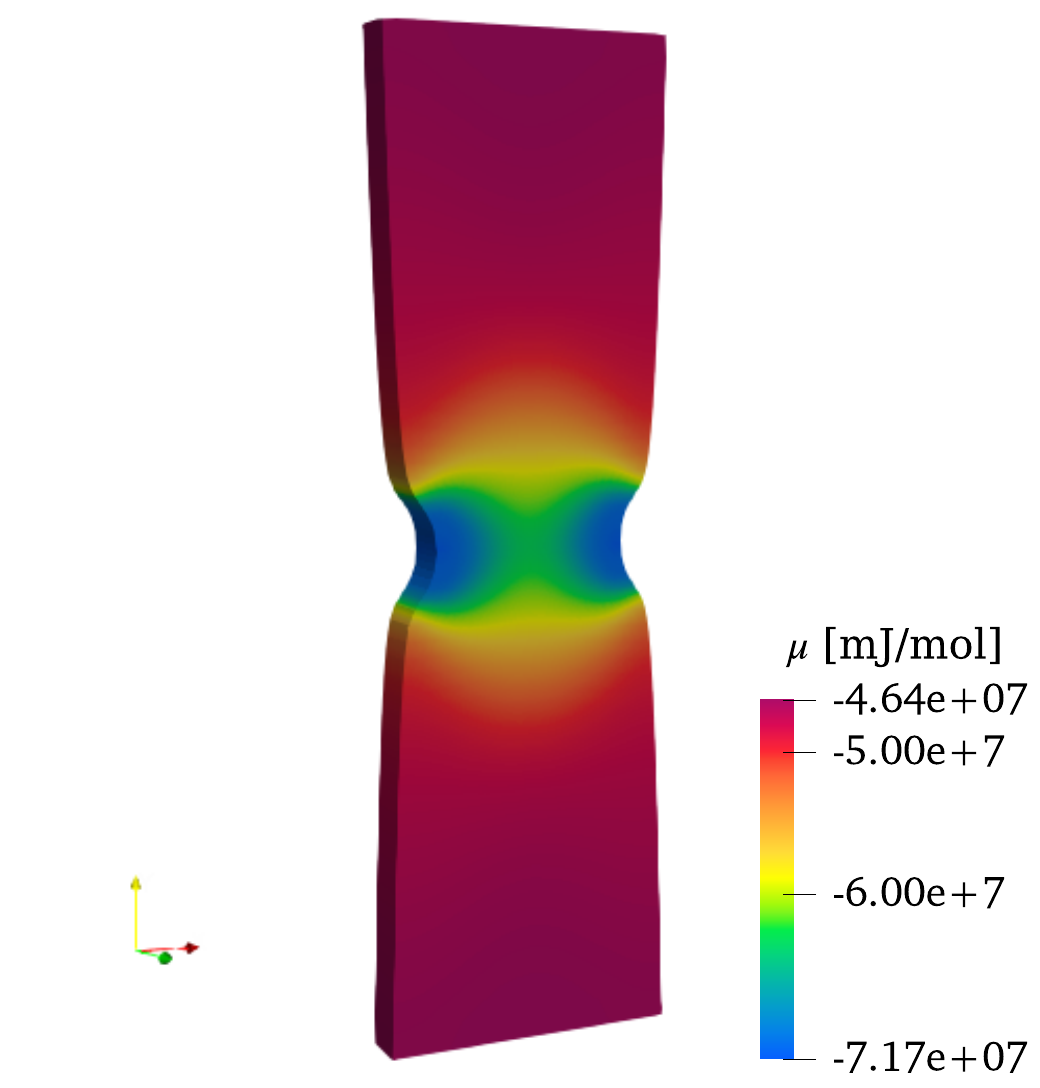}
	\includegraphics[width=0.45\textwidth]{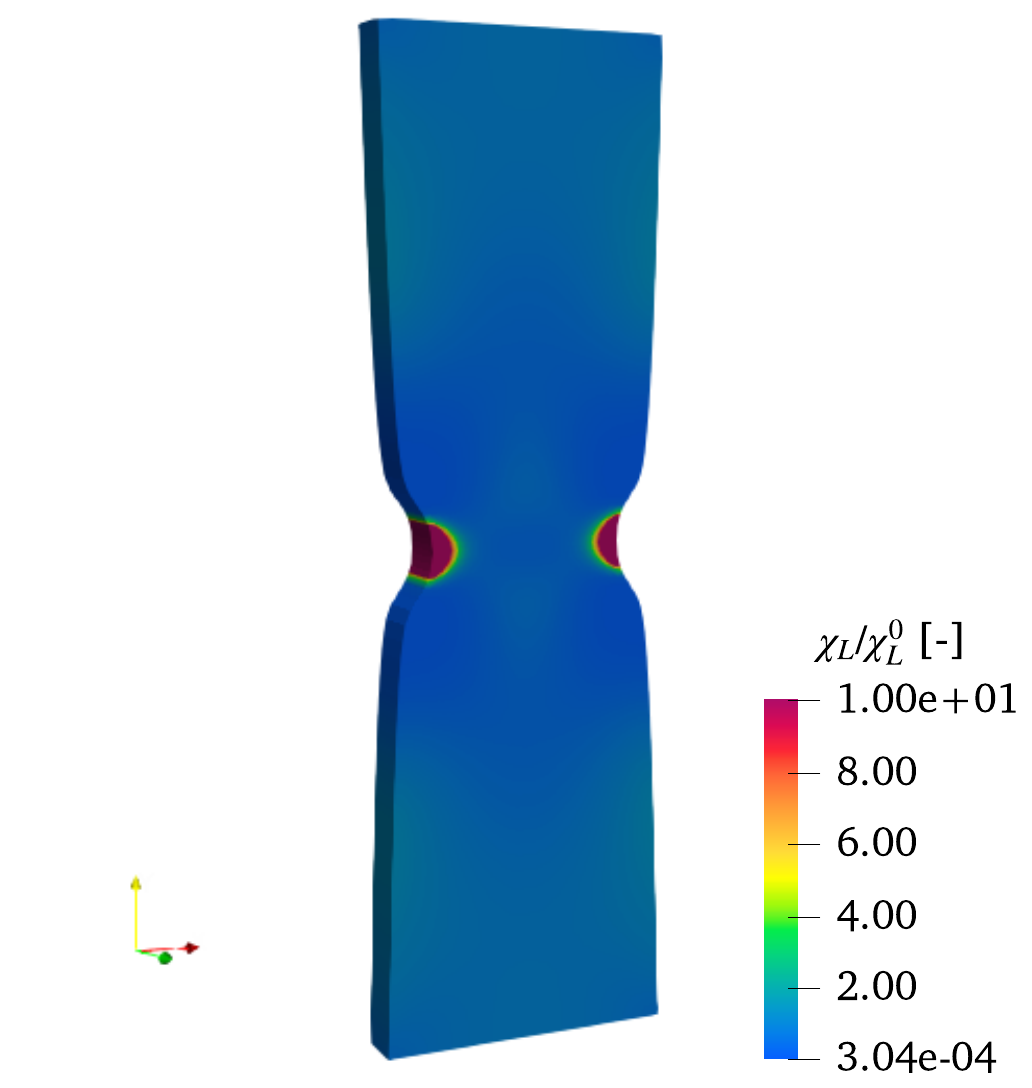}\\
	(a) Chemical potential [mJ/mol] $\hspace{1.0 cm}$ (b) $\chi_L/\chi_L^0$ [$-$]\\
	\includegraphics[width=0.45\textwidth]{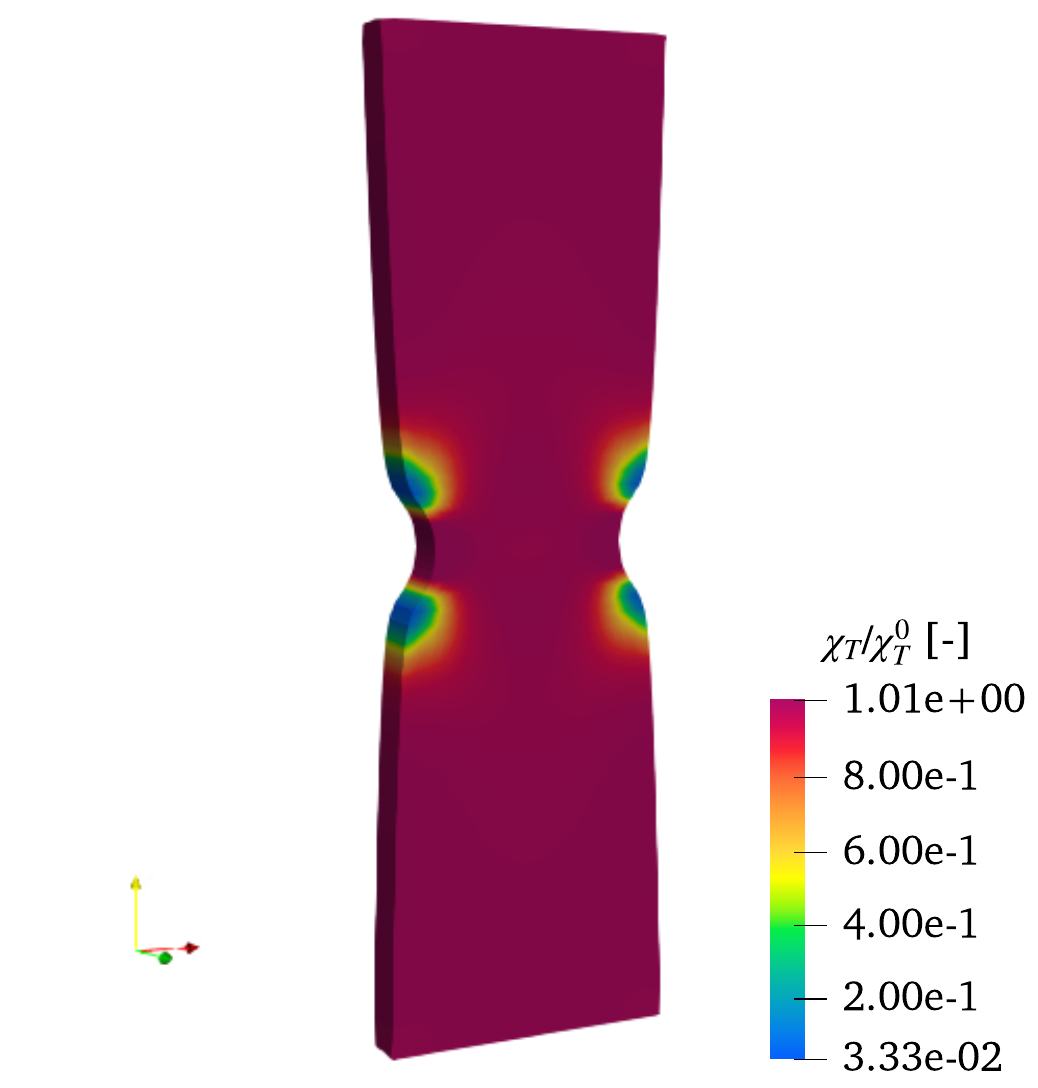}
	\includegraphics[width=0.45\textwidth]{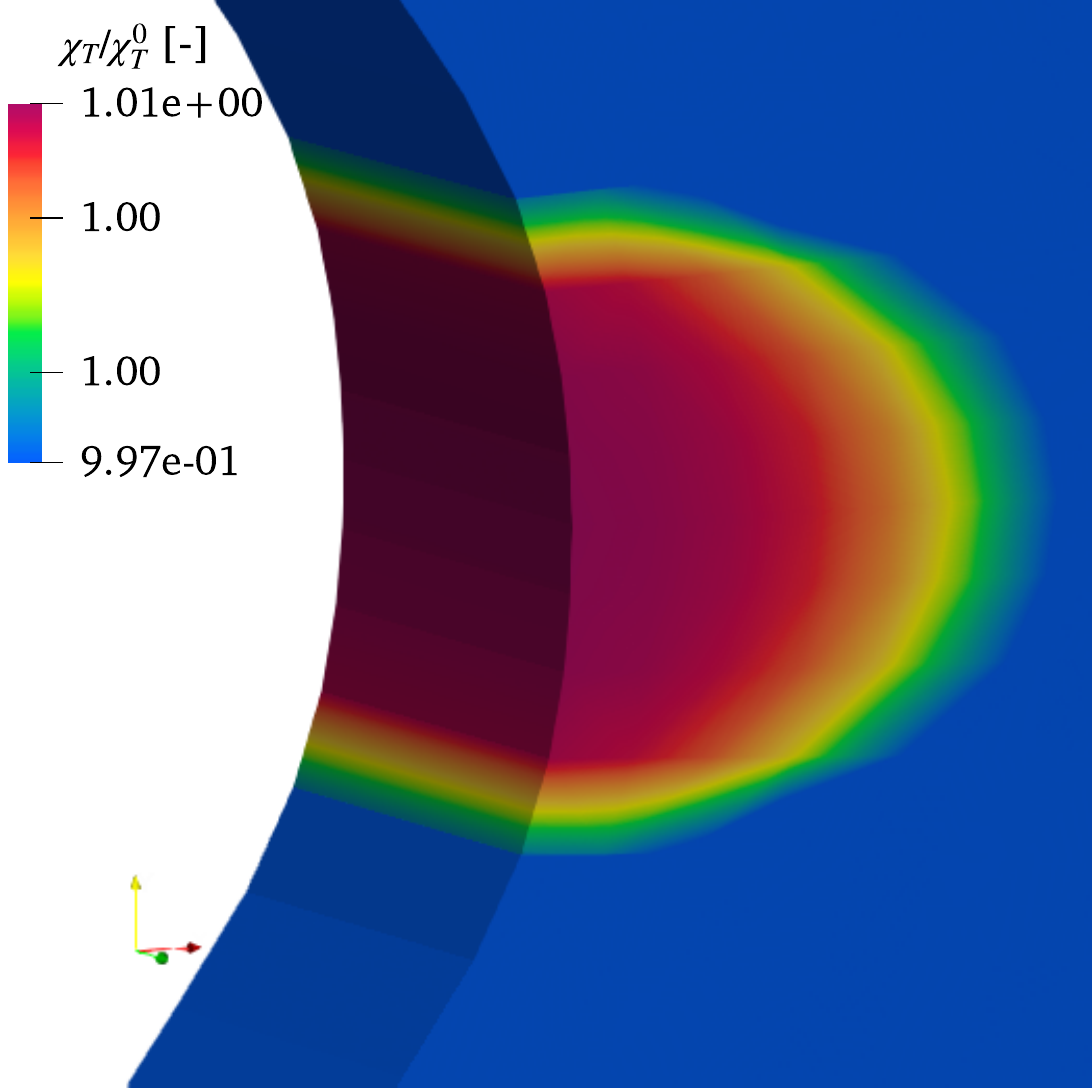}\\
	(c) $\chi_T/\chi_T^0$ [ $-$ ] $\hspace{2.0 cm}$ (d) Detail of $\chi_T/\chi_T^0$ [ $-$ ]\\
	\caption{Example~\ref{example-notched-specimen}: Chemical potential (a), normalized lattice hydrogen concentration (b), normalized trapped hydrogen concentration (c) and detail of normalized trapped hydrogen concentration (d) at time $t = 1$ s.}
	\label{fig_chi_notch}
\end{figure}

Figure~\ref{fig_chi_notch} shows the distributions of the chemical potential (a), and the normalized lattice (b) and trapped hydrogen concentration (c), at time $t = 1$ s. A detail of the normalized trapped hydrogen concentration (d) is also presented. As far as the chemical potential is concerned, note that we consider the case in which the boundary is chemically insulated. As a result of such insulation, the chemical potential drops in the notch since the hydrogen content in the domain must remain constant at its prescribed value. The analysis of the distributions of lattice and trapped hydrogen concentrations will be addressed in section~\ref{example-nt-pslip}.

\begin{figure}
	\centering
	\includegraphics[width=0.45\textwidth]{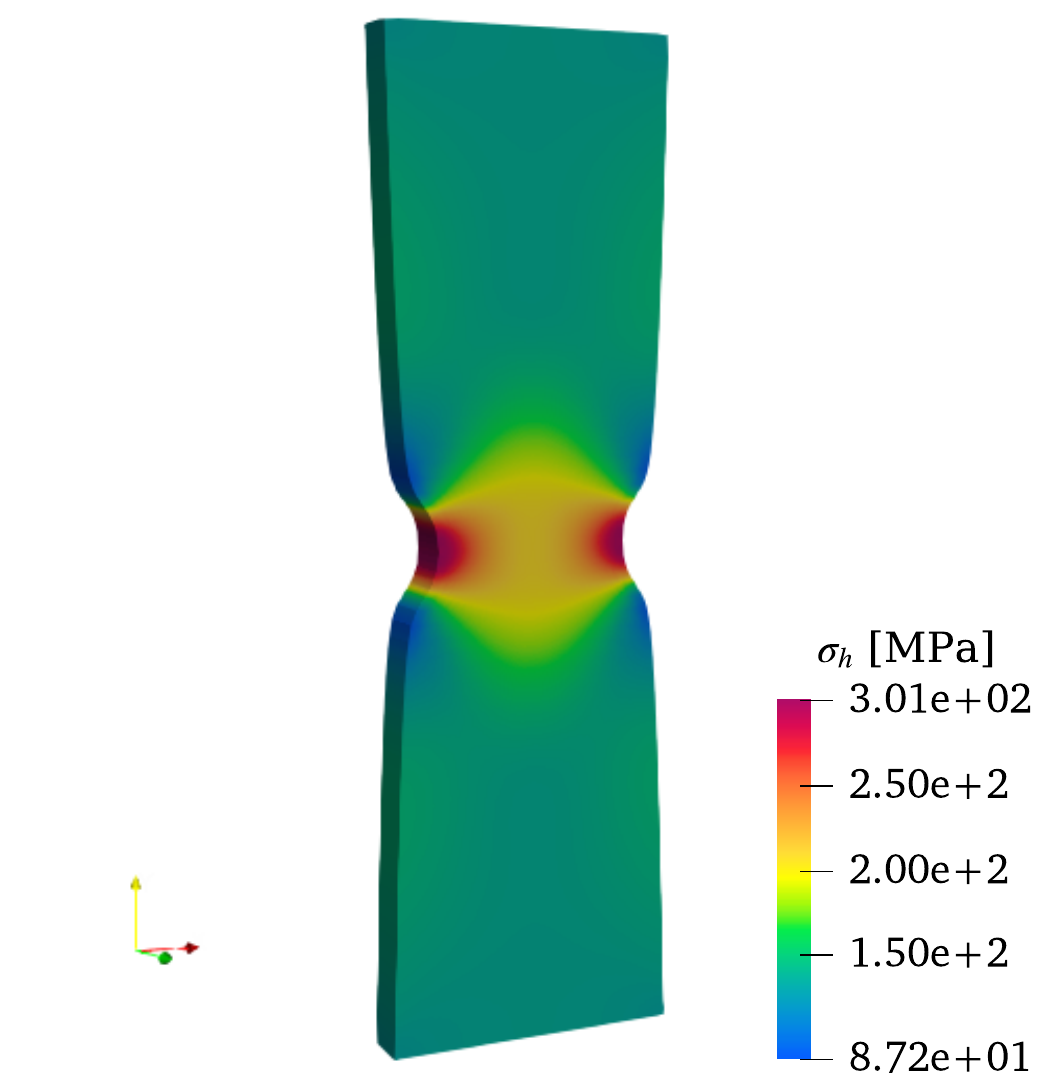}
	\includegraphics[width=0.45\textwidth]{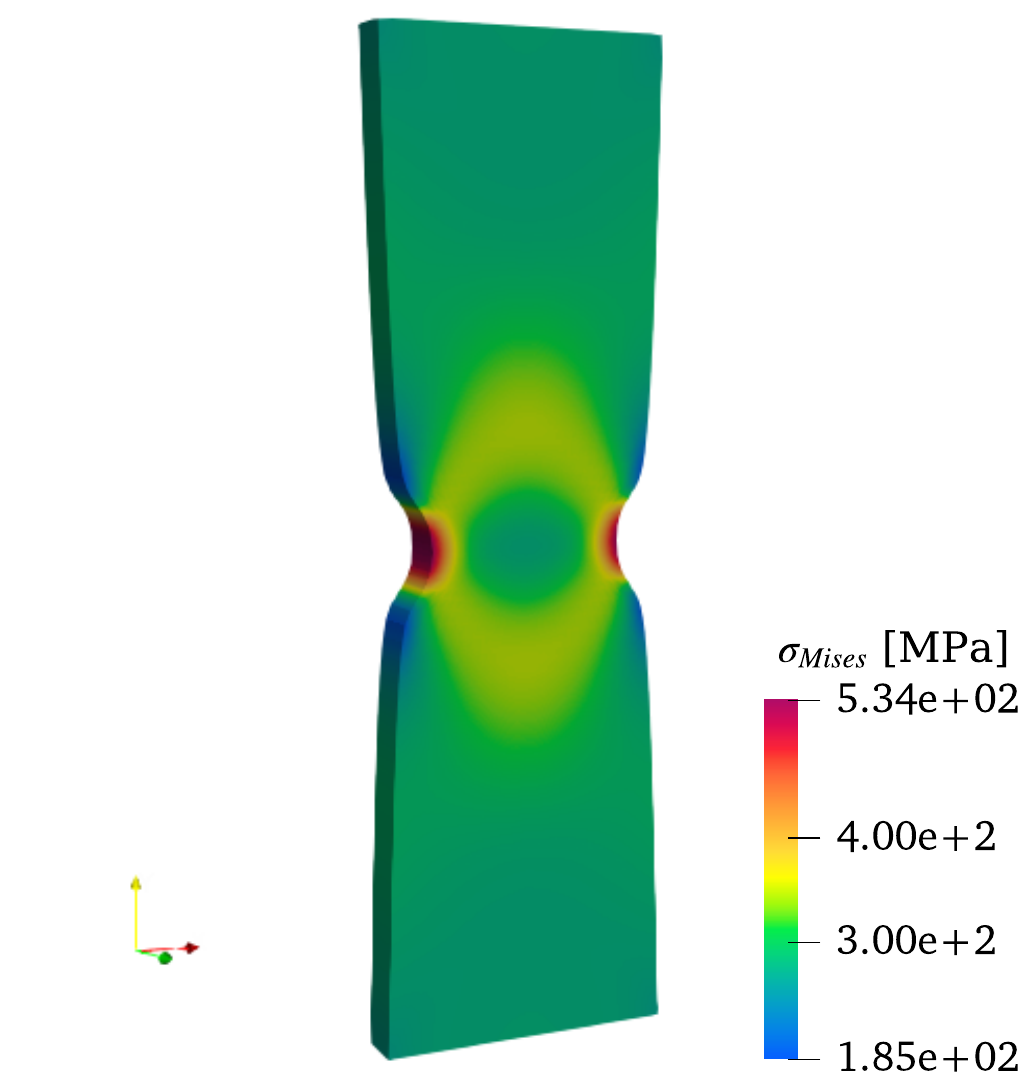}\\
	(a) Volumetric stress [MPa] $\hspace{2.0 cm}$ (b) Von Mises stress [MPa]\\
	\includegraphics[width=0.45\textwidth]{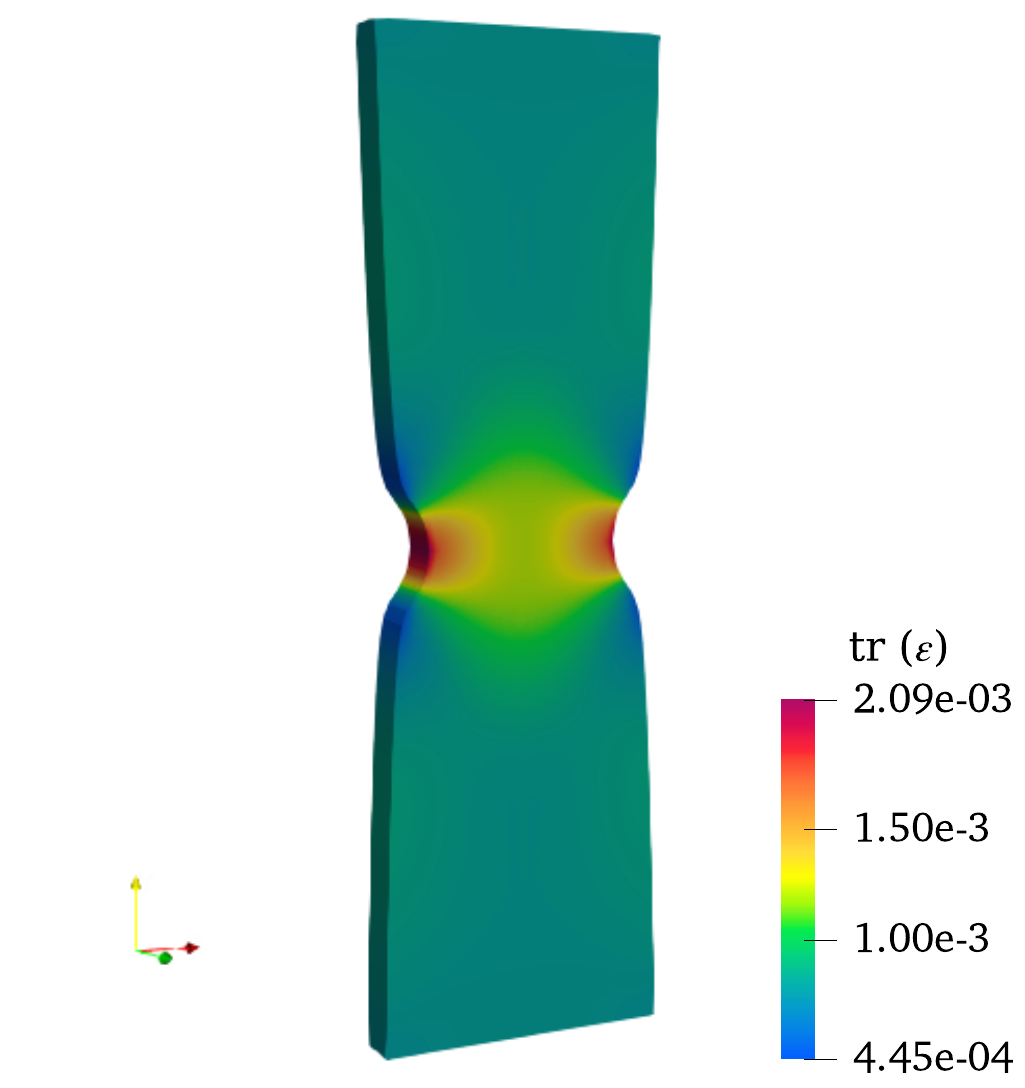}
	\includegraphics[width=0.45\textwidth]{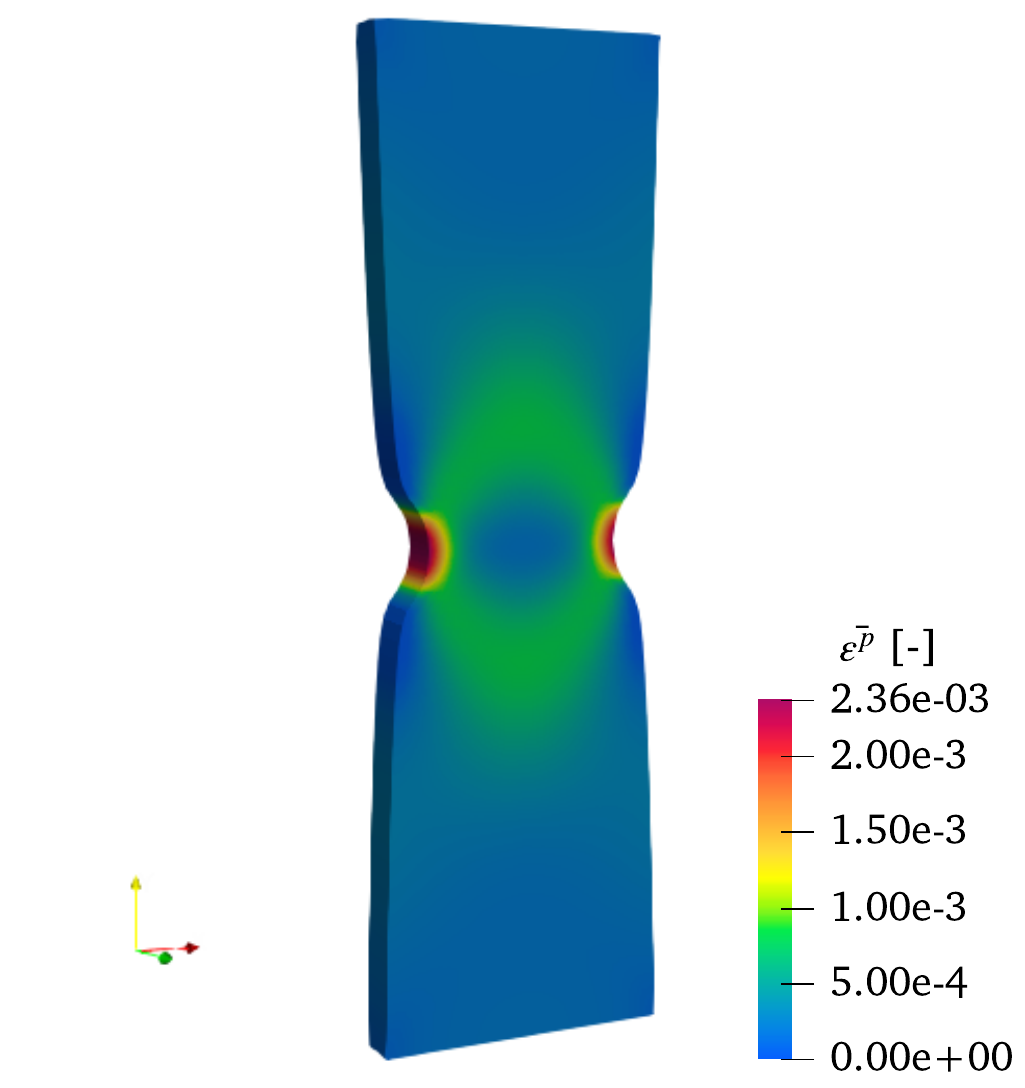}\\
	(c) Volumetric strain [-] $\hspace{3.5 cm}$ (d) Plastic slip [-]\\
	\includegraphics[width=0.45\textwidth]{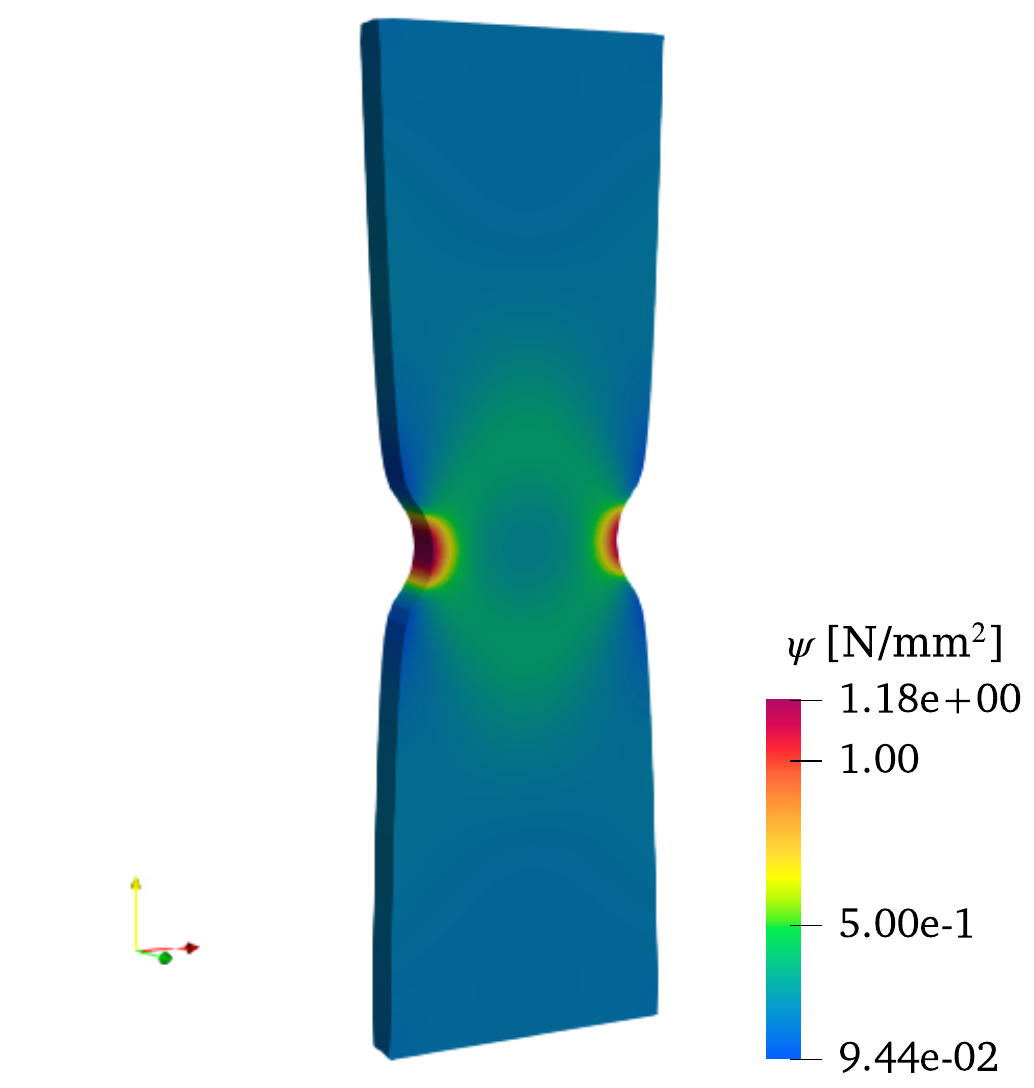}
	\includegraphics[width=0.45\textwidth]{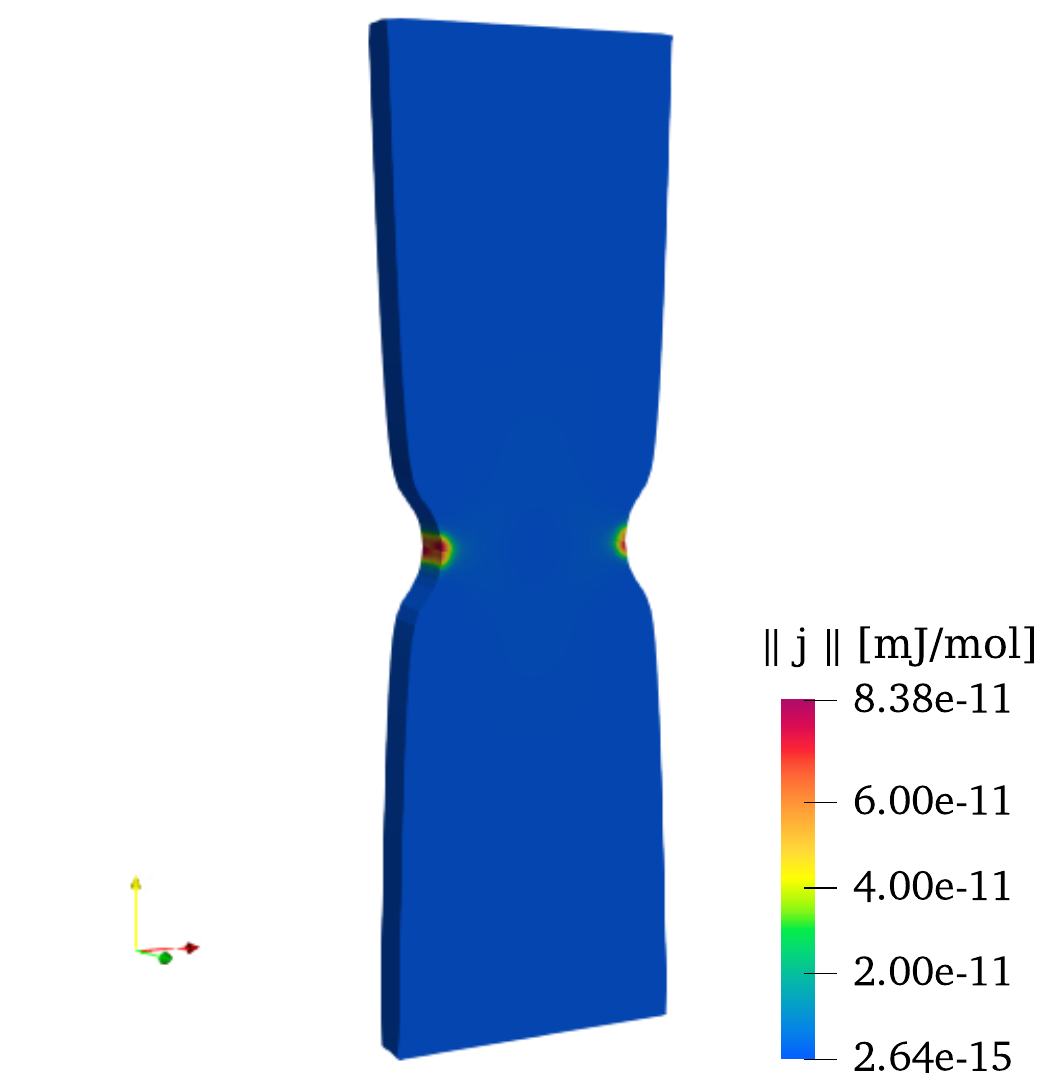}\\
	(e) Helmholtz free energy [N/mm$^2$] $\hspace{1.0 cm}$ (f) Mass flux norm [mJ/mol]\\
	\caption{Example~\ref{example-notched-specimen}: Volumetric stress (a), von Mises stress (b), volumetric strain (c), plastic slip (d), Helmholtz free energy (e) and norm of mass flux (f) at time $t = 1$ s.}
	\label{fig_mech_notch}
\end{figure}

Figure~\ref{fig_mech_notch} illustrates the fields of volumetric (a) and von Mises (b) stresses, the volumetric strain (c) and plastic slip (d), and, finally, the Helmholtz free energy (e) and the mass flux norm (f) distributions. We confirm that, qualitatively, they are in agreement with the expected results.  

\begin{figure}
  \centering
  \includegraphics[width=0.9\textwidth]{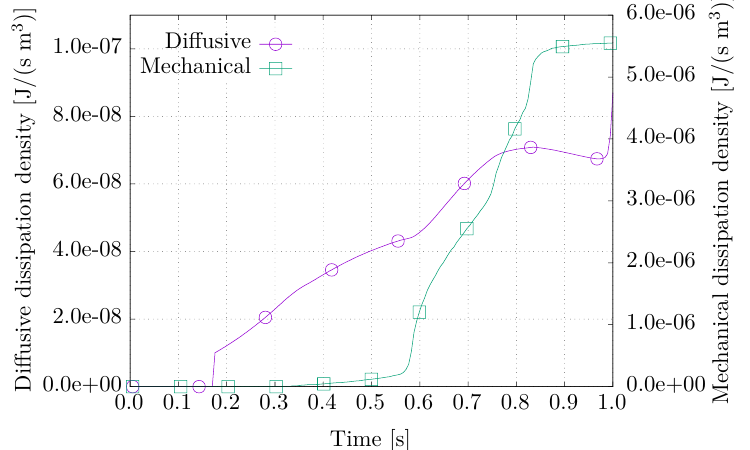}
	\caption{Example~\ref{example-notched-specimen}: Dissipation rate per unit volume: mechanical and diffusive contributions.}
	\label{fig_diss_notch}
\end{figure}

Figure~\ref{fig_diss_notch} depicts the mechanical and diffusive contributions to the dissipation during the simulation. Similarly to Example~\ref{example-plasticity}, the mechanical contribution starts to increase when plastic deformation is reached, and we confirm that both terms are non-negative.

\subsubsection{Effect of plastic deformation in the trapped hydrogen concentration $\chi_T$}
\label{example-nt-pslip}
To analyze the effect of the plastic deformation in the trapped hydrogen density $N_T$, and subsequently in the trapped hydrogen concentration $\chi_T$, we perform simulations employing the data fit proposed by Sofronis and McMeeking \cite{sofronis1989qb} (see Eq.~\ref{nt_expression_eq}), and without considering the evolution of $N_T$ with the equivalent plastic strain $\bar{\varepsilon}^p$, that is $N_{T2}$ = 0.

\begin{figure}
  \centering
  \includegraphics[width=1.0\textwidth]{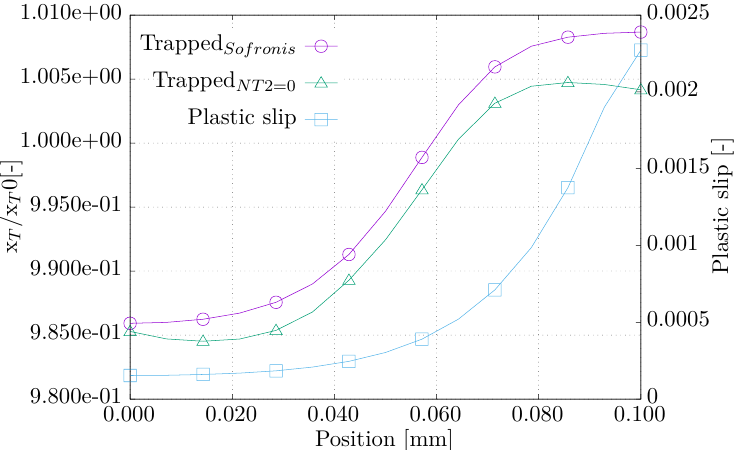}\\
  (a) Normalized trapped hydrogen concentration and plastic slip in the notch\\
  \includegraphics[width=0.78\textwidth]{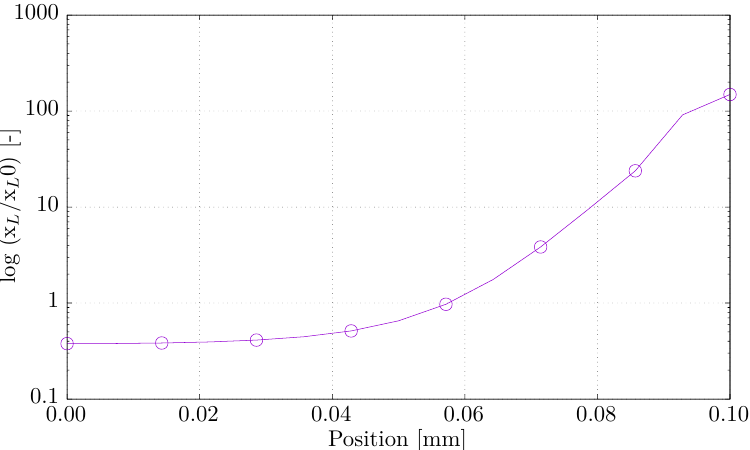}\\
	(b) Normalized lattice hydrogen concentration in the notch\\
  \caption{Example~\ref{example-notched-specimen}: Effect of plastic deformation in the normalized trapped hydrogen concentration $\chi_T/\chi_T^0$ along the notch.}
\label{fig_chi_eps_notch}
\end{figure}

Figure~\ref{fig_chi_eps_notch} (a) depicts the distribution of normalized trapped hydrogen concentration for both solutions and the plastic slip along the $x$-axis contained in the notch, whereas Figure~\ref{fig_chi_eps_notch} (b) shows the normalized lattice hydrogen concentration in the same region. The hydrogen distribution in the trap sites shows an increase at the notch in both solutions. This peak is significantly lower if we do not consider the mentioned dependency of the trapped hydrogen density with the equivalent plastic strain. This result evinces the importance of including an expression relating $N_T$ with $\varepsilon^p$.

As it is clear from Figure~\ref{fig_chi_eps_notch} (b), the results obtained  using a constant lattice chemical potential boundary condition differ significantly from the one which would be obtained using a constant lattice hydrogen concentration boundary condition. In the case of using a constant $\chi_L$ boundary condition, $\chi_L/\chi{_L^0} = 1$ at $x = 0$. Nevertheless, since we impose the boundary condition using the chemical potential, there is an increase in $\chi_L/\chi{_L^0}$. The evolution along the path is coherent with the solved problem. 

%% file: 5conclusions.tex
\section{Summary and main results}
\label{sec-conclusion}
This article describes a novel discretization for the numerical approximation of the (small strain) stress-diffusion problem that is customarily employed to model hydrogen diffusion in metals. The approximation is presented after reviewing the balance equations, the thermodynamics, and the kinetics of the coupled problem. 

This coupled problem has three characteristic features that affect the design of any numerical approximation. First, the mechanical and chemical problems are strongly coupled; second, inelastic behavior should be taken into account, particularly plasticity, in the mechanical problem since it modulates the diffusion parameters. Third, the transport problem involves the diffusion of \emph{two} species, representing respectively the interstitial and trapped hydrogen. These two forms of hydrogen are not independent but, on the contrary, must satisfy chemical equilibrium at all points and instants.

The numerical method proposed discretizes in space the problem using standard finite elements and introduces a new incremental variational principle whose optimization provides the time-marching solution, effectively introducing the equilibrium of the two diffusing species using a constraint that is later eliminated. Remarkably, the space/time discrete solution of all the fields of interest can be obtained from minimizing a single functional that depends only on the displacement field and a single scalar chemical potential. 

As a direct consequence of the variational character of the method, the tangent operator is symmetric and compact storage and fast linear solvers can be employed. Asymptotically, both the storage requirements and the linear solver times of the proposed method halve the costs of standard formulations. Numerical examples confirm these claims and the generality of the approach.

The ultimate goal of this work is to present a complete framework for analyzing bodies subject to mechanical loads and under the effects of dilute species. For many problems of interest, particularly the effect of hydrogen in metals, small strain kinematics provides a sufficiently accurate framework. In this article, we have restricted our theoretical and numerical analysis to this situation; however, the finite strain coupled problem has a similar structure to the one studied here, and the variational methods introduced can be extended to this regime by combining the new ideas of the current article and methods previously investigated by the authors~\cite{romero2021dd}.

The complete implementation of the material model is included in MUESLI~\cite{portillo2017ur}, the open-source material library developed by the authors and freely available at \texttt{https://ignromero@bitbucket.org/ignromero/muesli.git}.
